\newif\iflong
\newif\ifshort
\newif\ifaltermain
\altermaintrue

\iflong 
\else
\shorttrue
\fi

\documentclass[a4paper,11pt]{article}
\usepackage{fullpage}
\usepackage[square,numbers]{natbib}
\usepackage{authblk}

\usepackage{amsmath,amssymb,amsthm} %
\usepackage{paralist}
\usepackage{color}
\usepackage[colorinlistoftodos,bordercolor=orange,backgroundcolor=orange!20,linecolor=orange,textsize=scriptsize,disable]{todonotes}
\usepackage{soul}
\usepackage{mathtools}
\usepackage{dsfont}

\newtheorem{theorem}{Theorem}
\newtheorem{lemma}{Lemma}
\newtheorem{claim}{Claim}[theorem]

\newtheorem{proposition}{Proposition} 
\newtheorem{observation}{Observation}
\newtheorem{question}{Question}
\newtheorem{conjecture}{Conjecture}

\theoremstyle{definition}
\newtheorem{definition}{Definition}
\newtheorem{example}{Example}
\renewcommand{\emph}[1]{{\color{blue!60!black}\it{#1}}}
\usepackage{hyperref} 
\hypersetup{%
 backref=true, 
 pagebackref=true, 
 hypertexnames=true,
 colorlinks=true,citecolor=green!35!black,linkcolor=red!60!black%
}

\usepackage{cleveref}

\crefname{table}{Table}{Tables}
\crefname{figure}{Figure}{Figures}
\crefname{theorem}{Theorem}{Theorems}
\crefname{definition}{Definition}{Definitions}
\crefname{corollary}{Corollary}{Corollaries}
\crefname{observation}{Observation}{Observations}
\crefname{lemma}{Lemma}{Lemmas}
\crefname{example}{Example}{Examples}
\crefname{reduction}{Reduction}{Reductions}
\crefname{construction}{Construction}{Constructions}
\crefname{subsection}{Subsection}{Subsections}
\crefname{section}{Section}{Sections}
\crefname{proposition}{Proposition}{Propositions}
\crefname{algorithm}{Algorithm}{Algorithms}
\crefname{drule}{Rule}{Rules}
\crefname{claim}{Claim}{Claims}
\crefname{appendix}{Appendix}{Appendix}
\crefname{conjecture}{Conjecture}{Conjectures}

\usepackage{tikz}
\usetikzlibrary{decorations,arrows,petri,topaths,backgrounds,shapes,positioning,fit,calc,decorations.pathreplacing,patterns,intersections,decorations.pathmorphing,matrix}

\tikzset{linemarkr/.style =   {%
    opacity=.4,
    line width= 7pt,
    draw=Wcolor}} 
\tikzset{linemarkg/.style =   {decorate, decoration={snake,amplitude=.55mm,segment length=3mm}, line width= 2.5pt, draw=Ucolor}}
\tikzset{linemarkb/.style =   {line cap=round, opacity=.15, line width= 7pt, blue}}
\tikzset{linemarky/.style =   {line cap=round, opacity=.2, line width= 7pt, yellow}}

\tikzset{smalllinemarkstyle/.style={#1,opacity=.2, line width = 7pt}}
\tikzset{bpstyle/.style={#1,opacity=.5, line width = 4pt}}
\tikzset{agentnode/.style={draw, circle, fill=white, minimum size=1ex, inner sep=0pt, fill=black}}
\tikzset{labelnode/.style={inner sep=1.5pt, font=\small}}
\tikzset{cutsetcircle/.style={draw, gray, opacity=0.8, line width=1.3pt, dashed}}
\tikzset{extranode/.style={draw=blue!80!black, circle, fill=white, minimum size=1ex, inner sep=0pt, fill=blue!70!black}}

\usepackage{etoolbox} %
\newcommand{\tocommentout}[1]{%
}

\newcommand{\toappendix}[1]{%
  \gappto{\appendixtext}{
    {#1}
   }
}

\newcommand{\toappendixproofcontinued}[4]{%
  #1
   \gappto{\appendixtext}{
    \subsection{Continuation of the proof of \cref{#2}}\label{proof:#2}%
    \noindent{\normalfont\emph{#3}}

    {#4}
    }
}

\newcommand{\appendixproofwithstatement}[3]{%
  \gappto{\appendixtext}{
    \subsection{Proof of \cref{#1}}\label{proof:#1}
    \noindent {\normalfont\emph{#2}}
    #3
  }
}

\newcommand{\appendixcorrectnessproofwithstatement}[4]{%
  #1  
  \gappto{\appendixtext}{
    \subsection{Correctness of the Construction in the Proof
      of \cref{#2}}\label{proof:#2}
    {\normalfont\emph{#3}}

    #4
    }
}

\newcommand{\appendixsection}[1]{%
  \gappto{\appendixtext}{
    \section{Additional Material for Section~\ref{#1}}
    \label{appsec:#1}
  }
}

\renewcommand{\toappendixproofcontinued}[4]{%
  {#4}
}

\renewcommand{\toappendix}[1]{%
  {#1}
}

\renewcommand{\appendixproofwithstatement}[3]{%
  #3
}

\renewcommand{\appendixcorrectnessproofwithstatement}[4]{%
  #4
}

\usepackage{xspace}

\newcommand{\gencutset}{\text{cutset}\xspace}

\newcommand{\cutset}{\text{elementary cutset}\xspace}
\newcommand{\efouter}{\text{\normalfont{EF1\textsubscript{outer}}}\xspace}

\newcommand{\efone}{\text{\normalfont{EF1}\xspace}}
\newcommand{\eefone}{\emph{EF1}\xspace}
\newcommand{\todoHinline}[1]{\todo[inline, bordercolor=orange, linecolor=yellow!70!black, backgroundcolor=yellow!10]{H: #1}}

\newcommand{\todoBinline}[1]{\todo[inline, bordercolor=green!40!black, linecolor=green!70!black, backgroundcolor=green!10]{B: #1}}
\newcommand{\newH}[1]{#1}
\newcommand{\oldH}[1]{}
\usepackage{bibentry}

\newcommand{\mytitle}{Cutsets and EF1 Fair Division of Graphs}

\title{\mytitle}

\author[1]{Jiehua Chen} %
\author[2]{William S. Zwicker}
\affil[1]{TU Vienna, Austria\\
  jiehua.chen@ac.tuwien.ac.at}
\affil[2]{Murat Sertel Center for Advanced Economic Studies, Bilgi University,  Turkey\\
  and  Union College, Schenectady, New York, USA\\
  zwickerw@union.edu}
\date{}

\newcommand{\BibTeX}{\rm B\kern-.05em{\sc i\kern-.025em b}\kern-.08em\TeX}

\begin{document}

\pagestyle{plain}

\maketitle

\begin{abstract}
  In fair division of a connected graph~$G = (V,E)$,
  each of $n$ agents receives a share of $G$'s vertex set $V$. These shares partition $V$, with each share required to induce a \emph{connected} subgraph. Agents use their own valuation functions to determine the non-negative numerical values of the shares, which determine whether the allocation is fair in some specified sense. We introduce forbidden substructures called \emph{graph cutsets},
which block divisions that are fair in the \efone\ (envy-free up to one item) sense by cutting the graph into ``too many pieces''.
  Two parameters -- \emph{gap} and \emph{valence} -- determine blocked values of $n$.
  If $G$ guarantees connected \efone\ allocations for $n$ agents with valuations that are \emph{CA} (common and additive),
  then  $G$ contains no \emph{\cutset} of gap $k \geq 2$ and valence in the interval $[n-k + 1, n-1]$. 
  If $G$ guarantees connected \efone\ allocations for $n$ agents with valuations in the broader \emph{CM} (common and monotone) class, then
  $G$ contains no \gencutset\ of gap $k \geq 2$ and valence in the interval $[n-k + 1, n-1]$.
  These results rule out the existence of connected \efone\ allocations in a variety of situations.
 For some graphs $G$ we can, with help from some new positive results, pin down $G$'s \emph{spectrum} -- the list of exactly which values of $n$ do/ do not guarantee connected EF1 allocations. Examples suggest a conjectured common spectral pattern for all graphs.
   Further, we show that it is NP-hard to determine whether a graph admits a \gencutset.
We also provide an example of a (non-traceable) graph on eight vertices that has no cutsets of gap $\geq 2$ at all, yet fails to guarantee connected \efone\ allocations for three agents with \emph{CA} preferences. 
\end{abstract}

\section{Introduction}
In the original, continuous setting for fair division~\cite{Cakecutting96,Cakecutting,Ariel2016}, a single divisible good or ``cake,'' often
modeled by the closed interval~$[0, 1]$, is divided into $n$ pieces, with each agent allocated
a different piece of the resulting partition. One thread of this literature studies
allocations that are both envy-free (each agent values her assigned piece at least as highly as she
values any of the other pieces) and connected (each piece forms a single subinterval of~$[0, 1]$).

For the alternative setting of indivisible items, a finite set~$O$ of indivisible goods is partitioned into
disjoint subsets, with each agent allocated a different subset from the partition.
This context precludes envy-freeness as a
reasonable goal; for example, if $O$ contains but a single item, only one agent can get it. 
Budish~\cite{Budish} proposed a relaxation, \emph{envy-freeness up to one good} (in short \emph{\efone}), that circumvents this obstacle. It requires that whenever one agent $i$ envies %
another agent $j$, there exists some item in $j$'s share whose removal would eliminate that envy.
An \efone\ allocation always exists and it can be found via a simple envy cycle elimination algorithm in polynomial time~\cite{LMMS2004}. 

Fair division of graphs, our context here, provides a natural way to import the connectivity requirement from the continuous world into the world of indivisible goods.
The vertices of a finite connected graph~$G = (V, E)$ are viewed as indivisible
items, and we insist that the share of vertices allocated to each agent 
 form a connected subgraph.  Natural applications represented by this model (and mentioned in the paper~\cite{IgZwi} of \citeauthor{IgZwi}) include, for example, the problem of dividing cities connected by a road network among several parties, as when an island is partitioned and each
party wishes to drive among its allocated cities without leaving its own territory.
Alternatively, consider offices allocated to several departments of an organization, where
an edge represents a section of corridor joining a pair of offices in the organization's building, and each department
must receive contiguous offices.
It is known that connectivity and \efone\ can be incompatible. Graph I  of Figure 1 provides a particularly simple example and is discussed in \cref{sec:preliminaries}.   

  There are other cases that do guarantee \efone\ allocations.
  For a traceable graph (one that admits a Hamiltonian path) and up to four agents, \citeauthor{Bilo}~\cite{Bilo} provide an algorithm to construct a connected \efouter\ allocation even if the four agents have arbitrary but monotone valuations.
  Here, \efouter\ means that no agent shall envy another agent after an \emph{outer} item is removed from the share of the second agent, and an item is called \emph{outer} if removing it does not destroy connectivity.
  Igarashi~\cite{Igarashi} recently extended the positive result to arbitrarily many agents: %

 \begin{proposition}[\cite{Bilo}, \cite{Igarashi}]\label{IgThm}
   For each traceable graph~$G$ and positive integer~$n \geq 1$, and for all monotone valuations of the $n$ agents, there exists a connected \emph{\efouter} allocation. 
\end{proposition}

\noindent We restate this result informally, as follows:
Under arbitrary monotone valuations, traceable graphs \emph{universally} guarantee connected \efouter\ allocations.
Here \emph{universally} conveys that the result holds for arbitrarily many agents. %

Is the positive result behind \cref{IgThm} balanced by a corresponding negative one, showing that every non-traceable graph fails this universal guarantee? This remains the most important open question in the study of \efone\ graph fair division.

\begin{question} \label{Q1}
  Do any \emph{non-traceable} graphs offer
  the same universal guarantee? %
  Does the answer change if the valuation functions are CA? 
\end{question}

Note that for continuous graphs, however, such a general negative result \emph{does} exist, as shown by \citet{IgZwi}.
The continuous analogue of traceable is called \emph{stringable}, and stringability characterizes the class of continuous graphs that universally guarantee existence of connected EF allocations.\footnote{\label{tangle}A continuous graph, aka a \emph{tangle}, is a topological space wherein each edge of some connected graph is replaced by a copy of the $[0,1]$ interval of real numbers. Full EF replaces \efone\ in this context, and \emph{stringable} tangles correspond to traceable graphs. The characterization result for tangles uses a type of tangle cutset that is related to, but distinct from (and with a much simpler definition than) our graph cutsets here.} %

\smallskip
\noindent \textbf{Our contributions.}
While the discrete context of connected graphs still lacks a corresponding negative result of full generality, the \emph{graph cutset{s}} we introduce here represent progress to that end.
The cutset itself is a set of subgraphs; when the vertices in those subgraphs are excised, $G$ falls into a number of disconnected sections. Some agent $j$ will wind up with a share~$A_j$ that lacks enough critical vertices from the deleted subgraphs to form a path connecting any pair of disconnected sections.
This confines $A_j$ to a single section, so that if one has chosen the valuations appropriately, then $A_j$'s value is too small to be fair, in the \efone\ sense. 
In \cref{thm:main}, a main contribution, we make this idea precise, showing that graph cutsets constitute obstructions to connected \efone\ allocations.

 By \citet{Igarashi}, graph cutsets also obstruct traceability of a graph.  %
Suppose we think of a graph that has such an obstruction as being ``clearly non-traceable.''
Then our results here can be rephrased as follows: Every clearly non-traceable graph fails to universally guarantee connected \efone\ allocations.
Some non-traceable graphs fail to be ``clearly'' non-traceable, as we show in \cref{JCScounterexample}, but it seems possible that further generalizations of the cutset concept might address this shortfall.

Cutsets generalize one direction of the characterization, in Bil\'{o} et al.~\cite{Bilo}, of graphs that guarantee connected \efone\ allocations for $n = 2$ agents as those containing no \emph{trident} (see \cref{TridentDef}),
regardless of whether valuations are common and additive (CA) or common and monotone (CM). 
Our results suggest that the {CA} vs.\ {CM} distinction may first become  consequential for \efone\ graph division when there are more than $2$ agents (see \cref{sec:preliminaries} for the definitions).

 In addition to discussing universal results (the partial answer to Question 1),
 we also describe two special restrictions on graphs that guarantee the existence of  \efouter\ allocations once the number~$n$ of agents becomes large enough. 
 From algorithmic and complexity points of view, we show that \gencutset{s} are NP-hard to detect. We also analyze the existence of \efone\ allocations for some interesting graphs and conjecture that the guarantee of existence displays a certain common pattern for all graphs, as the number of agents varies. %

\smallskip
\noindent \textbf{Related work.}
Since its introduction~\cite{BG2015,Bout}, fair division of graphs has been among the most relevant research topics of fair division with constraints. 
Recent work in this setting investigates different fairness concepts with respect to matters of existence and (parameterized) complexity~\cite{Bilo,IgZwi,Igarashi,Igarashi_Peters_2019,BILS,TruLonc2020,GHS2020,MSVV2021,parameterizedFair,DeligkasGraph}.

\newH{The graph cutsets we introduce here are 
related to cutsets for \emph{tangles} (``continuous graphs''; see \cref{tangle}), introduced by \citet{IgZwi}, but with some differences. Envy-free fair division of tangles tells us a lot about \efone\ fair division of \emph{saturated} graphs (in which each edge has at least one endpoint of degree one or two), but less about non-saturated ones.  Differences between the graph and tangle definitions of cutset seem driven, in part, by additional subtlety in defining graph cutsets for non-saturated graphs. The graphical cake model in \cite{Elkind} is similar to the tangle concept, but cannot be used to draw direct conclusions about graphs because with graphical cake, agents' shares are allowed to overlap at the vertices.
We refer the reader to 
the survey paper by Suksompong~\cite{Suksompong2021} on fair division with other constraints.
}

\smallskip

\noindent \textbf{Paper outline.}
{The rest of the paper is organized as follows.
In \cref{sec:preliminaries}, we introduce necessary definitions and concepts for connected \efone\ division of graphs.
In \cref{sec:cutsets}, we define our main concept of ``graph cutsets'' and provide two main obstruction results: \begin{inparaenum}
  \item The existence of a \emph{tame} graph cutset precludes the graph from guaranteeing \efone\ allocations even when all agents have common and additive (CA) valuations.
  \item  The existence of a graph cutset precludes the graph from guaranteeing \efone\ allocations even when all agents have common and monotone (CM) valuations.
\end{inparaenum} 
We also show a counterexample to any converse, in the form of a graph~$G$ containing no cutsets of any kind, yet there exist CA valuations for $3$ agents that rule out the existence of a connected \emph{\efone} allocation.  %
In \cref{sec:cutset:NP-hard}, we show that finding a graph cutset is indeed NP-hard.
In \cref{sec:spectrum}, we analyze how different graphs behave in terms of \efone\ allocations for increasing numbers of agents.
\newH{In particular, we show two positive cases that guarantee the existence of connected \efouter\ allocations.}
We conclude with some future research directions. }

\section{Preliminaries: EF1 Divisions of Graphs}\label{sec:preliminaries}
\appendixsection{sec:preliminaries}
Let $N=\{1,2, \dots, n\}$ be a finite set of \emph{agents} and $G=(V,E)$ be a {connected} undirected graph.
We call $G$ \emph{traceable} if it admits a Hamiltonian path.
We call a vertex subset $V'\subseteq V$ \emph{connected} or a (connected) \emph{piece} if it induces a connected subgraph of $G$.
Each agent $i \in N$ has a \emph{valuation} -- a function $v_i\colon 2^{V} \rightarrow \mathds{R}^+$ assigning non-negative real values to  connected pieces, with $v_i(\emptyset)=0$.
A valuation~$v_i$ is \emph{monotone} if for all $X,Y\in \mathcal{C}(V)$ it holds that $X \subseteq Y$ implies $v_i(X) \leq v_i(Y)$.
Monotone valuations treat vertices as \emph{goods}; we do not consider bads (or chores) here.
The valuation functions of the agents are called \emph{common} if $v_i = v_j$ holds for all $i,j\in N$, and are \emph{arbitrary} if not required to be common. Valuations are \emph{additive} if $v_i(I) = \sum_{x\in I} v_i(\{x\})$ for each agent $i$ and each piece $I\in \mathcal{C}(V)$.  We will use abbreviations \emph{CM} for ``common and monotone'', and \emph{CA} for ``common and additive.'' Additive valuations form a proper sub-class of monotone valuations (because of the non-negativity constraint on valuations), and CA forms a proper sub-class of CM. 
A (connected) \emph{allocation}~$A = \{ A_i \}_{i \in N}$
of $G$ 
assigns each agent $i\in N$ a connected piece $A_i \in \mathcal{C}(V)$, with these pieces \emph{partitioning} $V$, so that $\bigcup_{i\in N} A_i = V$ and $A_i \cap A_j = \emptyset$ when $i\neq j$. %

In fair division of a graph $G$, we ask whether there exists such an allocation that is \emph{fair}, in some well-defined sense.  \emph{Maximin share fairness} was the principal fairness criterion studied by Bouveret et al.~\cite{Bout}, which first introduced the topic of graph fair division.   Three later works -- \cite{Bilo}, \cite{Igarashi}, and \cite{IgZwi} --  instead focus (as does this paper) on the two variants of \emph{envy-freeness} defined below.
The original definition of \emph{envy-freeness} requires, of an allocation $A$ that   $v_i(A_i) \ge v_i(A_j)$ hold for every pair $i,j\in N$ of agents%
\iflong
\else{. }
\fi 
With indivisible objects, envy-free allocations may not exist, and so we instead use the following notions.  

\begin{definition}[\cite{Budish}]\label{def:EF1:outer}
  An allocation $A=(A_1,A_2,\dots,A_n)$ of the vertices in a graph~$G$ is \emph{envy-free up to one good}, aka \emph{\efone}, if for each pair $i,j$ of agents, either $v_i(A_{i}) \geq v_i(A_{j})$, or there is an element~$x$ of $A_j$ such that $v_i(A_i) \geq v_i(A_j \setminus \{ x \})$.
  $A$ is \emph{envy-free up to one outer good}, aka \emph{\efouter} if for each pair $i,j$ of agents, either $v_i(A_{i}) \geq v_i(A_{j})$, or there is an element  $x$ of $A_j$ such that $A_j \setminus \{ x \}$ is connected in~$G$ and $v_i(A_i) \geq v_i(A_j \setminus \{ x \})$.
\end{definition}

 Two previous results concern which untraceable graphs guarantee existence of an \efouter\ allocation  for~$n = 2$ or $n=3$ agents. The first is a complete characterization for the two-agent case; the second applies to three agents, but is narrower in scope. The first result requires two additional notions:

\begin{definition}[Bipolar orderings and tridents] \label{TridentDef}
  Let $G=(V,E)$ denote a connected graph. 
  A \emph{bipolar ordering} of a graph $G$ is a linear order~$x_1, x_2$, $\dots , x_k$ of $G$'s vertices such that every initial segment $x_1, x_2, \dots , x_m$ ($1 \leq m \leq k$) of the order induces a connected subgraph, as does every final segment $x_m, x_{m+1}, \dots , x_k$.
  Equivalently, each vertex $x_i$ is adjacent to some vertex $x_j$ appearing earlier on the order (unless $i = 1$) and is adjacent to some vertex $x_j$ appearing later on the order (unless $i = k$).

  Let $C \subseteq V$ be a subset of vertices, and let $G \setminus C$ denote the subgraph of $G$ induced by the vertex set $V \setminus C$.  
  \begin{compactenum}[(1)]
    \item\label{def:type-I-trident} If $|C|=1$, and $G \setminus C$ has three or more connected components, then $C$ is a \emph{type-1 trident}. 
    \item\label{def:type-II-trident} If $|C| > 1$; $G\setminus C$ has exactly three connected components $H_1, H_2,$ and $H_3$; for each $H_j$  exactly one vertex $s_j \in C$ (referred to as $H_j$'s \emph{contact vertex}), is adjacent to any vertices of $H_j$; and the vertices $s_1, s_2, s_3$ are distinct, then $C$ is a \emph{type-2 trident}.  
  \end{compactenum}
\end{definition}

Note that every Hamiltonian path \emph{is} a bipolar ordering, but the converse fails: Graph IV of \cref{fig:gap2-examples} is not traceable, but has a bipolar ordering -- for example, order the vertices from left to right, with the middle pair of vertices ordered either way.
Removing a trident cuts the graph into $3$ or more disconnected pieces.

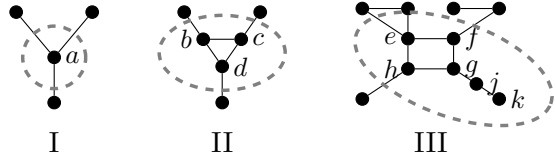
\begin{figure}[t!]
  \centering
  \def \xw {1}
  \def \yw {.6}
  \begin{tikzpicture}
    \foreach \x / \y / \n  in
    {0/1/1, 1/1/2, 0.5/0/3,
      0.5/-1/4} {
      \node[agentnode] at (\x*\xw, \y*\yw) (v\n) {};
    }

    \foreach \s / \t in {1/3, 2/3, 3/4} {
      \draw (v\s) -- (v\t);
    }
    
    \foreach \n / \la / \posx / \dis in {3/a/right/0} {
      \node[labelnode, \posx = \dis pt of v\n] {$\la$};
    }

    \draw[cutsetcircle] (v3) circle[radius=2.5ex];

    \node[below = 1ex of v4] {I};
  \end{tikzpicture}\qquad
  \begin{tikzpicture}
    \foreach \x / \y / \n  in
    {0/1/1, 1/1/2, 0.5/-0.2/3, 
      0.5/-1/6} {
      \node[agentnode] at (\x*\xw, \y*\yw) (v\n) {};
    }
    
    \path (v3) -- node[agentnode,pos=0.5] (v4) {} (v1);
    \path (v3) -- node[agentnode,pos=0.5] (v5) {} (v2);

    \foreach \s / \t in {1/4, 3/4, 3/5, 2/5, 4/5,  3/6} {
      \draw (v\s) -- (v\t);
    }
    
    \foreach \n / \la / \posx / \dis in {4/b/left/0, 5/c/right/0, 3/d/right/0} {
      \node[labelnode, \posx = \dis pt of v\n] {$\la$};
    }

    \path (v4) -- node[midway] (midF) {} (v3); 
    \path (v5) -- node[midway] (midL) {} (v3); 
    \path (midF) -- node[midway] (mid) {} (midL); 
    \draw[cutsetcircle]  (mid) ellipse (5ex and 3ex);

    \node[below = 1ex of v6] {II};
  \end{tikzpicture}\qquad
  \begin{tikzpicture}
    \def \xw {.6}
    \def \yw {.4}
    \foreach \x / \y / \n  in
    {-1.5/2/1, -.5/2/2, .5/2/3, 1.5/2/4, 
     -0.5/1/5, 0.5/1/6, -0.5/0/7, 0.5/0/8,
     -1.5/-1/9, 1.5/-1/10%
   } {
      \node[agentnode] at (\x*\xw, \y*\yw) (v\n) {};
    }

    \foreach \s / \t in {1/2, 3/4, 1/5, 2/5, 4/6, 5/6, 5/7, 6/8, 7/8,7/9,8/10} {
      \draw (v\s) -- (v\t);
    }

    \path (v8)  -- node[agentnode,midway] (v11) {} (v10);

    \foreach \n / \la / \posx / \dis in {8/g/right/0, 11/j/right/0,
      10/k/right/0, 5/e/left/0, 6/f/right/0, 7/h/left/-1} {
      \node[labelnode, \posx = \dis pt of v\n] {$\la$};
    }

    \draw[cutsetcircle,rotate=-20]  (v8) ellipse (8.2ex and 3.8ex);

    \path  (v9) -- node[midway] (midF) {} (v10); 
    \node[below = 1ex of midF] {III};
    
  \end{tikzpicture}
  \caption{Graph I has a type-1 trident; Graphs~II and III have type-2 tridents.}
  \label{tridents}
\end{figure}
\begin{example}\label{ex:tridents}
   \emph{Here we review material found in \cite{Bilo}, explaining how tridents act to block 
 \efone  fair divisions for $n = 2$ agents.} \cref{tridents} shows three connected graphs;
  Graph I contains a type-1 trident, while Graphs II and III each contain a type-2 trident (consisting of all vertices with labels).
  In Graph~I, deleting vertex~$a$ would disconnect the graph into three components (each of which consists of a single vertex).
  Only one agent can receive a piece including $a$; we say that this agent \emph{dominates}~$a$.
  The piece of the second agent (we call him \emph{deprived}) contains at most
  a single one of the three components, as he cannot use~$a$ to forge a connected share from vertices belonging to different components. Next, assume that each of the four vertices has value 1 to both agents, and the value of a set of vertices is obtained by summing the values of the individual vertices. Then the deprived agent receives at most single vertex of value 1, while the other agent receives at least three vertices, each worth 1 to the deprived agent, thus leaving him envious of the other agent by more than one item. In other examples, deleting vertex $a$ might leave more than three connected components, each with more than one vertex. We can similarly force envy by assigning value $1$ to $a$ and to one vertex from each component, with value $0$ assigned to all other vertices.  In these situations, vertex $a$ is an example of a \emph{type-1 trident} -- an obstacle to connected \emph{\efone}%
  allocations for two agents, when paired with a suitable choice of CA valuations.
  Similar arguments, deferred to \cref{ex:tridents}, apply to  Graphs II and III.

Deleting any single vertex from Graph~II 
yields at most two disconnected components, so this graph has no type-1 trident.  It does have what we will call a \emph{type-2 trident}, however, in the form of the central subgraph $\mathcal{C}_{\text{II}}$ induced by vertex set $\{ b,c,d \}$, which acts collectively in a manner similar to a type-1 trident. Deleting the vertices in $\mathcal{C}_{\text{II}}$ would yield three disconnected components, and only one agent can \emph{dominate} $\mathcal{C}_{\text{II}}$ by being allocated at least two of the three vertices  $b,c,d $.  The share of a second, \emph{deprived} agent contains at most one of these three vertices -- not enough for him to form a connected share containing vertices from more than one of the components. Note that this last argument requires that each component have its own distinct \emph{contact point} $s \in \mathcal{C}$, with $s$ adjacent to a vertex in that component. 
For a type-2 trident, the counterexample CA valuations that defy connected \emph{\efone} allocations for $2$ agents are a bit different.  We assign value $\frac{1}{3}$ 
to each of the three contact points $ b,c$ and $d $ of $\mathcal{C}_{\text{II}}$, value $1$ to one vertex from each of the three components, and value $0$ to all other vertices (from the components, or from $\mathcal{C}_{\text{II}}$).\footnote{Value $1$ would also work for the contact vertices here, but related examples require a strictly smaller value.}  The deprived agent now receives a share with at most two valuable vertices, whose values are $1$ and $\frac{1}{3}$, while the agent who dominates $\mathcal{C}_{\text{II}}$ %
is left with at least four valuable vertices with values $1$, $1$, $\frac{1}{3}$, and $\frac{1}{3}$, so he is envied by more than one item.

 What  about variants of Graph II for which deleting a subgraph $\mathcal{C}$ leaves more than three components?  This  happens in  Graph III if we declare the trident to be the central square %
 induced by vertices $e,f,g,$ and $h$. We would then get 4 components, each with its own distinct contact point in %
 the square. However, we disallow tridents with more than three contact points (and they are not needed for the 2-agent characterization).  In the case of Graph III we can instead enlarge the square by having our trident $\mathcal{C}_{\text{III}}$ absorb the fourth component (with vertices $ j$ and $k$ in \cref{tridents}) completely, as suggested by the dashed gray ellipse in the figure. For the $2$ agent case, the same can be done any time the proposed type-2 trident has more than $ 3$ contact points, or has $3$ contact points with multiple components sharing a common contact point.  
\end{example}

We are ready to state the two known results.
\begin{proposition}[\cite{Bilo}] \label{Bilo2Agents}The following are equivalent for all finite connected graphs~$G$:
\begin{inparaenum}[(i)]
    \item $G$ guarantees connected \emph{\efouter} allocations for $2$ agents with arbitrary monotone %
    {valuations.}
    \item $G$ guarantees connected \emph{\efouter} allocations for $2$ agents with CA valuations.
    \item $G$ contains no tridents.
    \item $G$ has a bipolar ordering.
  \end{inparaenum}
\end{proposition}

\begin{proposition}[\cite{IgZwi}]
  \label{LipsThm}  The lips graph,\footnote{The lips graph of \citeauthor{IgZwi} has vertices $a, b,$ and $c$; two edges join $a$ to $b$, two join $b$ to $c$, and one joins $a$ to $c$. Graph $L^\star$ of Figure 3 shows a version with added subdivision vertices $v_1$ -- $v_5$.  } and all of its subdivisions, guarantees connected \emph{\efouter} allocations for $3$ agents with monotone valuations.    
\end{proposition}

The proof of \cref{LipsThm} uses a discretization of a modified version of Stromquist's famous moving knife argument~\cite{Stromquist} for continuous fair division of the $[0,1]$ interval. The technique works for a few other graphs, 
but we know of no characterization for $3$ agents analogous to \cref{Bilo2Agents}. For \emph{positive} results, the situation for $4$ or more agents is worse yet -- there are none, except those for traceable graphs already implied by \cref{IgThm}, and those implied, for a few very small graphs, by two additional special cases that we provide    %
in \cref{sec:spectrum}.

The graph cutsets we introduce in \cref{def:non-generalized-cutset} provide new negative results for a variety of specific graphs and values of $n \geq 3$, thus generalizing the role of tridents in \cref{Bilo2Agents}, which applied only to $n=2$ agents. A cutset $\mathcal{C}$ resembles a set of tridents that achieve the necessary disconnections by acting collectively. If any of these are type-II, then $\mathcal{C}$ is a \emph{generalized cutset} (\gencutset, for short), if at most one is type-II then $\mathcal{C}$ is a \emph{tame generalized cutset} (tame \gencutset, for short), and if none are type-II then $\mathcal{C}$ is an \emph{elementary cutset.}
\section{Graph Cutsets and Main Obstruction Theorem}\label{sec:cutsets}
\begin{figure}[t!]
  \def \xw {.6}
  \def \yw {.6}
  \def \rscale {.8}
  \centering
  \begin{tikzpicture}
    \foreach \x / \y / \n / \la / \posx / \dis in
    {0/0/1/e/left/0, .9/0/2/a/right/0, 2/1/3/c/right/0, 2/-1/4/d/right/0, 3.1/0/5/b/left/0, 4/0/6/f/right/0} {
      \node[agentnode] at (\x*\xw, \y*\yw) (v\n) {};
    }

    \foreach \x / \y / \n / \la / \posx / \dis in {1/0/2/a/right/0, 3/0/5/b/left/0} {
      \node[labelnode, \posx = \dis pt of v\n] {$\la$};
    }
    \foreach \s / \t in {1/2, 2/3, 2/4, 3/5, 4/5, 5/6} {
      \draw (v\s) -- (v\t);
    }

    \draw[cutsetcircle] (v2) circle[radius=3ex*\rscale];
    \draw[cutsetcircle] (v5) circle[radius=3ex*\rscale];

    \node[below = 1ex of v4] {IV};
  \end{tikzpicture}~~
  \begin{tikzpicture}
    \foreach \x / \y / \n / \la / \posx / \dis in {0/0/1/e/left/0, .9/0/2/a/right/0, 2/1/3/c/right/0, 2/-1/4/d/right/0, 3.1/0/5/b/left/0, 4/0/6/f/right/0} {
      \node[agentnode] at (\x*\xw, \y*\yw) (v\n) {};

    }
    \foreach \s / \t / \n / \la in {2/3/7/d, 2/4/8/e} {
      \draw (v\s) -- node[agentnode, pos=.4] (v\n) {} (v\t);
    }
    \foreach \n / \la / \posx / \dis in {2/c/above/0, 7/d/right/1, 8/e/right/1, 5/f/left/0} {
      \node[labelnode, \posx = \dis*\xw pt of v\n] {$\la$};
    }

    \foreach \s / \t in {1/2, 3/5, 2/3, 2/4, 4/5, 5/6, 7/8} {
      \draw (v\s) -- (v\t);
    }

    \path (v7) -- node[midway] (mid) {} (v8); 
    \path (v2) -- node[pos=.8] (midX) {} (mid); 
    \begin{pgfonlayer}{background}
      \draw[cutsetcircle]  (midX) ellipse (2ex and 4ex);
    
      \draw[cutsetcircle] (v5) circle[radius=3ex*\rscale];
    \end{pgfonlayer}
    \node[below = 1ex of v4] {V};
  \end{tikzpicture}~~
  \begin{tikzpicture}
    \foreach \x / \y / \n / \la / \posx / \dis in {0/0/1/e/left/0, .9/0/2/a/right/0, 2/1/3/c/right/0, 2/-1/4/d/right/0, 3.1/0/5/b/left/0, 4/0/6/f/right/0} {
      \node[agentnode] at (\x*\xw, \y*\yw) (v\n) {};

    }
    \foreach \s / \t / \n  in {2/3/7, 2/4/8,5/3/9,5/4/10} {
      \draw (v\s) -- node[agentnode, pos=.4] (v\n) {} (v\t);
    }

    \foreach \n / \la / \posx / \dis in {2/g/above/0, 7/h/right/1, 8/j/right/1, 5/m/above/0, 9/k/left/0, 10/n/left/0, 1/a/below/0, 6/b/below/0,
    3/c/right/0, 4/d/right/0} {
      \node[labelnode, \posx = \dis*\xw pt of v\n] {$\la$};
    }

    \foreach \s / \t in {1/2, 3/5, 2/3, 2/4, 4/5, 5/6, 7/8, 9/10} {
      \draw (v\s) -- (v\t);
    }

    \begin{pgfonlayer}{background}
      \path (v7) -- node[midway] (mid) {} (v8); 
      \path (v2) -- node[pos=.8] (midX) {} (mid); 
      \draw[cutsetcircle]  (midX) ellipse (2ex and 3.8ex);
      
      \path (v9) -- node[midway] (mid) {} (v10); 
      \path (mid) -- node[pos=.8] (midX) {} (v5); 
      \draw[cutsetcircle]  (midX) ellipse (2ex and 3.8ex);
    \end{pgfonlayer}

    \node[below = 1ex of v4] {VI};
  \end{tikzpicture} 
  \caption{Three examples of cutsets with valence 2 and gap $\ge 2$,
    forbidding connected \efone\ allocations for $3$ agents under CA valuations.}\label{fig:gap2-examples}
\end{figure}
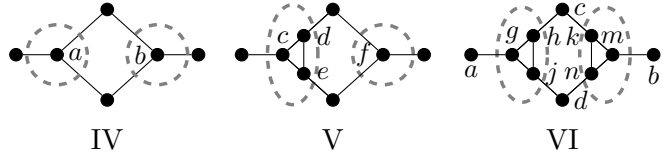

\appendixsection{sec:cutsets}
Cutsets provide obstacles to connected \emph{\efone} allocations for more than just two agents, generalizing tridents; see \cref{TridentDef}.
The consecutive numbers $1,2$ and $3$ played a critical role in a type-1 trident; %
we removed $1$ point from a connected graph, we had $2$ agents, and the point's removal yielded $3$ subgraphs that were disconnected from one another.  Suppose instead we remove $2$ points from the graph, we have $3$ agents, and when we remove both of the points we get $4$ disconnected subgraphs?  This is exactly the situation for Graph IV in Figure~\ref{fig:gap2-examples}, when removing the circled points $a$ and $b$. It will follow from \cref{def:generalized-cutset-final} that $\mathcal{C}_{\text{IV}} =  \{  a  , b  \}$ is an \emph{elementary cutset} of \emph{gap}~$\geq 2$.
The word ``gap'' here refers to the difference between the number of points removed and the number of disconnected subgraphs that result. We see next how connected \emph{\efone} allocations for three agents are blocked by a cutset like the one for Graph~IV.

\newcommand{\noinstances}{%
  For each graph in \cref{fig:gap2-examples}, there exist CA valuations for $n=3$ agents such that no connected \efone\ allocations exist (so no connected \efouter\ allocations exist).
}

\begin{lemma}%
  \label{lem:no-instances}
  \noinstances
\end{lemma}

\begin{proof}
  We consider each graph one by one.
  For graph~IV, given any partition of the vertices into three connected shares, at most one agent dominates vertex $a$ (meaning $a$ is in her share) and at most one other dominates $b$.  With three agents, this leaves some third agent (we call him \emph{deprived}) dominating neither member of the cutset. Just as in the earlier argument for Graph I (with the type-I trident) his share will contain vertices from at most one of the four disconnected subgraphs. Consider the CA valuation that assigns value~$1$ to $a$, $1$ to $b$, $1$ to exactly one vertex selected from each of the four disconnected subgraphs, and value $0$ to all remaining vertices (if there are any -- of course, there are none for Graph~IV). The deprived agent gets at most one valuable vertex, worth~$1$, leaving $5$ valuable vertices, worth $1$ each, to be split between the remaining two agents.  One agent gets at least three of these, and the deprived agent envies her by more than one item.

  Graph V is different; removing two vertices from the graph never yields more than \emph{three} disconnected subgraphs.\footnote{It also holds for traceable graphs that deleting $k$ vertices yields no more than $k+1$ disconnected subgraphs.} But the subgraph induced by $c,d,$ and $e$ acts like  a type-2 trident;  at most one agent $x$ can \emph{dominate} the subgraph (meaning $x$'s share contains at least two of the three contact points $c,d,$ and $e$).  We set the gap $\geq 2$ cutset to be $\mathcal{C}_{\text{V}} = \{ \{ f \}, \{ c,d,e \}  \}$, our first example of a \emph{generalized cutset}; see \cref{def:non-generalized-cutset}. The additional set braces clarify that $\mathcal{C}_{\text{V}}$ has two members -- a \emph{type-I member} resembling a type-1 trident, and a \emph{type-II member} resembling a type-2 trident. Deleting the vertices from both members yields $4$ disconnected subgraphs, while the number of agents who can dominate at least one cutset member is at most $2$, making the gap equal to $4-2=2$.  We can think of each cutset member as being a kind of ticket, on which a \emph{face value} of $1$ is printed: 

 {\centering \it 1 pass-through allowed. \par}

 \noindent Any agent who dominates a member of $\mathcal{C}_{\text{V}}$ gets to use the ticket, granting her the possibility to obtain a share containing vertices from two of the four subgraphs.
 Yet other examples have tickets with face values $k > 1$ as well; these can be used by $k$ agents and correspond to cutset members with enough contact points to be passed through by $k$ agents.
 See Graph~{VII} from \cref{fig:VII+JCS+Lips-graph} and \cref{lem:graphVII} for more discussion.
 
The \emph{valence} of a cutset is the sum of the face values on the tickets, so $\mathcal{C}_{\text{V}}$ has valence $2$, as does cutset $\mathcal{C}_{\text{IV}}$ for Graph~IV.
When the number of agents is greater than the valence of the cutset (at least three, for graphs IV or V), we know that some \emph{deprived} agent has no ticket to use, so his share has vertices from at most one of the disconnected subgraphs; when the number of agents is less than the number of components (at most three, for graphs IV or V), some (other) \emph{privileged} agent must have a share overlapping two disconnected subgraphs. %

We use $\mathcal{C}_{\text{V}}$ to block connected \emph{\efouter} allocations for $3$~agents with CA valuations as follows: Assign weight $1$ to each of the four unlabeled vertices of Graph V -- that's one vertex from each disconnected subgraph, weight~$1$ to vertex~$f$, and weight $\frac{1}{3} $ to each of the vertices $c,d,e$; if there were any additional vertices, we would have assigned them weight $0$.  Now the deprived agent receives at most \emph{two} valuable vertices, with values of $1$ and $\frac{1}{3} $.  The remaining agents receive at least three unlabeled vertices, so some agent $y$ gets at least two of these.  These two come from different disconnected subgraphs, so to connect them agent $y$ must also get a ``ticket,'' meaning $y$'s share either includes $f$ or  includes at least two of the three vertices in $\{c,d,e\}$.  So agent $y$ either gets three vertices valued at $1$, $1$, and $1$, or gets four vertices valued at $1$, $1$, $\frac{1}{3} $ and $\frac{1}{3} $.  Either way, the deprived agent envies $y$ by more than one item. 

\toappendixproofcontinued{}{lem:no-instances}{\noinstances}
{
The proof for Graph~VI is a bit more involved.
  Consider the cutset~$\mathcal{C}_{\text{VI}} = \{ \{ g,h,j \} , \{ k,m,n \} \}$, with $2$ type-II members, and valence of $2$.
{
  We assign weight~$\frac{1}{3}$ to each of the vertices~$g$ and $m$, and weight~$\frac{1}{4}$ to each of the remaining vertices from the cutset, i.e., $h,j,k,$ and $n$.
  We assign weight~$1$ to each of the vertices ~$a,b,c,$ and $d$ not from the cutset~$\mathcal{C}_{\text{VI}}$.
  With three agents, there will again be a deprived agent Alice who gets at most one vertex from $\{g,h,j\}$, at most one vertex from $\{k,n,m\}$, and at most one vertex from~$\{a,b,c,d\}$.
  So Alice's share must be one of the following (or a subset of one):
  \begin{compactenum}[(1)]
    \item\label{alice-1}  $\{h, c, k\}$ (or $\{j, d, n\}$); the worth is $1/4 + 1 + 1/4$.
    \item\label{alice-2} $\{a, g\}$ (or $\{b,m\}$); the worth is $1 + 1/3$.
  \end{compactenum}
  Some privileged agent Lily gets at least two from $\{a,b,c,d\}$, hence also gets either at least two from $\{g,h,j\}$ or at least two from $\{k,n,m\}$.
  Thus Lily's share must be one of the following (or a superset of one):
  \begin{compactenum}[(1)]
    \setcounter{enumi}{2}
    \item\label{lily-1} $\{c,h,j,d\}$ (or $\{c,k,n,d\}$); the worth is $1 + 1/4 + 1/4 + 1$.
    \item\label{lily-2} $\{a,g,h,c\}$ (or $\{a,g,j,d\}$); the worth is $1 + 1/3 + 1/4 + 1$.
    \item\label{lily-3} $\{b,m,k,c\}$ (or $\{b,m,n,d\}$); the worth is $1 + 1/3 + 1/4 + 1$.
  \end{compactenum}
  Now Alice envies Lily by more than one item, except in one case: Alice's share is of type \eqref{alice-1} (and not a proper subset of type \eqref{alice-1}) and Lily's is of type \eqref{lily-1}. But this combination is impossible, as their shares would intersect.
}}
\end{proof}
\newH{\noindent The type of additive valuations used in the proof above for Graph V also work more generally for \cutset{s} that are \emph{tame}, meaning they are limited to at most a single type-II member, but this approach falls apart for the general case when there are two or more such members. For the particular case of Graph~VI we were only able to come up with CA valuations (that forbade connected \efone\ allocations for three agents) thanks to the final argument, establishing that for connected allocations the share of a deprived agent cannot contain single vertices from two different type-II members  without overlapping the share of a privileged agent.
There exist other graphs in which shares are not forced to intersect in this way, leaving us without an argument showing existence of a CA counterexample. For these examples, a more general CM construction still works; see \cref{thm:main}.}

\begin{figure}%
  \centering
  \begin{tikzpicture}
  \def \xw {.5}
    \foreach \r / \a / \n  in
    {3/180/1, 1/180/2, 1/108/3, 
      1/252/6, 1/324/5,1/36/4} {
      \node[agentnode] at (\a:\r*\xw) (v\n) {};
    }

    \foreach \s / \t / \aa in {1/2/0, 1/3/0, 1/4/-65, 1/5/65, 1/6/0,2/3/0, 3/4/0,4/5/0,5/6/0,6/2/0} {
      \path[draw] (v\s) edge[bend right=\aa] (v\t);
    }

    \foreach \s / \t / \aa in {1/2/0, 1/3/0, 1/4/-65, 1/5/65, 1/6/0} {
      \path (v\s) edge[bend right=\aa] node[extranode,pos=0.45] {} (v\t);
    }

   \foreach \r / \a / \n  in
    {3.4/180/a, .6/180/b, .6/108/c,       .6/252/f, .6/324/e, .6/36/d} {
      \node at (\a:\r*\xw) (v\n) {$\n$};
    }

  \end{tikzpicture}~~
  \begin{tikzpicture}
  \def \xw {.55}
  \def \yw {.75}
    \foreach \x / \y / \n  in
    {-0.6/0/3, -1.6/-0.4/2, -2/0.5/1, 0.6/0/6, 1.6/-0.4/7, 2/0.5/8,
    0/.6/4, 0/-.6/5} {
      \node[agentnode] at (\x*\xw, \y*\yw) (v\n) {};
    }

    \foreach \s / \t / \aa in {1/2/0, 2/3/0, 3/4/0, 3/5/0, 4/6/0, 5/6/0, 6/7/0, 7/8/0} {
      \path[draw] (v\s) edge[bend right=\aa] (v\t);
    }

    \path[draw] (v2) edge[out=-50, in=-130] (v7);

    \foreach \n / \la / \posx / \dis in {1/a/left/0, 2/b/below left/-1,
    3/c/below/0, 4/d/below/0, 5/e/below/0, 6/f/below/0, 7/g/below right/-1, 8/h/right/0} {
      \node[labelnode, \posx = \dis pt of v\n] {$\la$};
    }

    \foreach \n / \la / \posx / \dis in {1/2/above/0, 2/2/above/0,
    3/2/above/0, 4/3/above/0, 5/3/above/0, 6/2/above/0, 7/2/above/0, 8/2/above/0} {
      \node[labelnode, \posx = \dis pt of v\n, red!80!black] {$\la$};
    }
  \end{tikzpicture}~~\begin{tikzpicture}
    \def \xw {.5}
    \def \yw {.45}
    \foreach \x  / \y / \n  in
    {0/0/b, -2/0/a, 2/0/c,
      -1/0/2, 1/0/4, -1/1/1, 1/1/3, 0/-1/5} {
      \node[agentnode] at (\x*\xw, \y*\yw) (v\n) {};
    }

    \foreach \s / \t / \aa in {a/2/0, 2/b/0, b/4/0, 4/c/0,
      a/1/-30, 1/b/-30, b/3/-30, 3/c/-30, a/5/30, 5/c/30%
    } {
      \path[draw] (v\s) edge[bend right=\aa] (v\t);
    }

    \foreach \n / \la / \posx / \dis in
    {a/a/left/0, b/b/below/0,
      c/c/right/0,
      1/{v_1}/above/0, 2/{v_2}/below/0, 3/{v_3}/above/0, 4/{v_4}/below/0, 5/{v_5}/below/0} {
      \node[labelnode, \posx = \dis pt of v\n] {$\la$};
    }
  \end{tikzpicture}

  \caption{Left: Graph VII has a cutset of valence 3 and gap~$\ge 2$. Middle: The JCS graph. Right: The graph~$L^*$ from Figure~11 in the paper of Igarashi and Zwicker~\cite{IgZwi}.}\label{fig:VII+JCS+Lips-graph}
\end{figure}
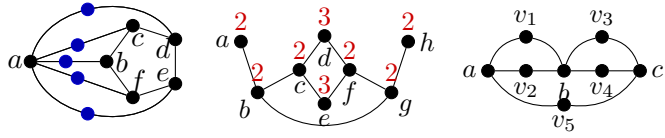
\begin{lemma}\label{lem:graphVII}
  Graph VII from \cref{fig:VII+JCS+Lips-graph} %
  has a cutset that blocks connected \emph{\efone} allocations for $4$ agents with  CA valuations.
\end{lemma}

\begin{proof}
  Our cutset for this graph is $\mathcal{C}_{\textbf{VII}} = \{ \{a\}, \{b,c,d,e,f \}  \}$, with two members; $H_1 = \{a\}$ is  type-I and $H_2 = \{b,c,d,e,f \}$ is type-II, but has $5$ contact points rather than only $3$.  Consider any connected share that contains $2$ of the $5$ unlabeled vertices of graph VII, but omits $a$.   To connect those two, it must include at least two of $H_2$'s vertices -- that is, it must dominate $H_2$.  But $H_2$ has only $5$ vertices, so it can be dominated by at most two agents.  Its ticket reads

{\centering  \it 2 pass-throughs allowed. \par}

\noindent This enables as many as two agents to pass through $H_2$, while an additional ticket enables one more agent to pass through $H_1$.  The valence of $\mathcal{C}_{\textbf{VII}}$ is the sum $1+2=3$ of the pass-through numbers on these two tickets.  The rest of the argument is similar to the one for graph~V.
With four agents, some deprived agent dominates neither member of $\mathcal{C}_{VII}$ and their share contains at most one unlabeled vertex. If we assign weight $1$ to $a$ and to each unlabeled vertex, and assign weight $\frac{1}{3}$ to each of the five contact vertices of $H_2$, then the deprived agent will envy one of the others by more than one item.  
\end{proof}

The seven examples from \cref{fig:gap2-examples,fig:VII+JCS+Lips-graph} portend all of the main ideas used in the general proofs of our main theorem, as well as most -- but not all -- of the features found in the precise definitions of \gencutset.  Other examples, not presented here, have stretched these definitions in two directions, relaxing some requirements so as to be satisfied by these other examples, while forcing the imposition of additional requirements not needed for graphs I -- VII.  We seek a definition as broad as possible, to encompass any situation for which a version of the deprived agent argument applies.  For example, we want to allow cutsets for which the ``gap'' (between valence of some cutset $\mathcal{C}$ and the number of connected components that remain after excising $\mathcal{C}$) is more than two, but the requirement that contact points be unique and distinct must also be imposed on these excess components, and these requirements then sometimes rule out cutset examples that we might think to include.  In some cases we can bypass that problem by merging several components into one.
\todoHinline{Need to rewrite the following for the extensions.}
The merged version is a subgraph that might no longer be internally connected, so we will not call it a ``component'' any more; note that the deprived agent argument only needs these subgraphs to be disconnected from each other, and never needs them to be internally connected as subgraphs.\footnote{Curiously, the argument also does not require that a type-II member be internally connected. But it is certainly easier to find cutsets hiding within a graph such that each type-II member is connected, and we show in Section \ref{appenxi:cust-relax} that imposing connectivity never limits the consequences for connected \emph{\efone} allocations.    
}  Other requirements are necessitated for generalized cutsets that are not tame (meaning they have more than one type-II member). For example, we need to be sure that the deprived agent cannot put together a path from one section to a different one by crossing from one type-II member to a different type-II member while using only one contact point from each.

\subsection{Main theorem}\label{sec:main}
\appendixsection{sec:main}
\ifaltermain

The definition given here for \cutset is quite simple -- much more so than that for the generalized version of \gencutset{s} to follow. An \cutset\ is just a set of $t$ vertices, that -- when excised -- break the graph into at least $t+2$ pieces that are disconnected from one another (expressed as ``at least $t+2$ connected components'' in the precise version, below).  The \emph{valence}, in this case, is just the number $t$ of vertices that are cut (but valence will become more complicated for \gencutset{s}).  This simpler type of cutset was exemplified in graphs I (with valence 1) and IV (with valence 2) of the previous section, and can be thought of as a generalization of the type-1 tridents from \citeauthor{Bilo}~\cite{Bilo}. We fold the argument showing that \cutset{s} can block the existence of connected \emph{\efone} allocations for agents with CA valuations into the more general version for tame cutsets given in our proof of \cref{thm:main}.\footnote{But note that a separate argument for this case would be shorter (and quite similar to what was seen in the examples).}

\begin{definition}[\cutset]\label{def:non-generalized-cutset}For $G = (V, E)$ a finite connected graph, let
$\mathcal{C}  = \{c_1, \dots, c_t\}$ be a set of $t$ vertices in $V$.
If the subgraph of $G$ induced by the vertex set $V \setminus \mathcal{C}$ contains at least $t+r$ nonempty connected components $H_1, \dots , H_t, H_{t+1}, \dots, H_{t+r}$ with $r\ge 2$, then $\mathcal{C}$ is a  \emph{\cutset} of \emph{gap} $r\ge 2$ and \emph{valence} $t$.
\end{definition}

In the definition above the valence reduces to being simply the number of the members of the cutset,  each of the sets in $H$ can be presumed to be connected, and there are no restrictions on connectivity between pairs of cutset members.  This now all changes with the more general definition. The valence of a cutset can now exceed its cardinality, as a type-II cutset member $C_j$ may have $2x + 1$ contact vertices, allowing as many as $x$ agents to \emph{dominate} $C_j$ by owning
shares that contain $2$ such vertices. Each of those $x$ agents could use these
contact vertices to connect a pair of vertices from different
components; Graph VII is an example.  Also, members of the partition $H$ may now be (internally) disconnected, and connections between pairs of type-II members are restricted.  These complications are needed, if we wish our main theorem to apply to the widest collection of cases. Before giving the definition, we need some additional terminology:

\begin{definition}[big $\bigcup$, independent family]\label{def:independent}
  For $X$ any family of subsets of a universe, let \emph{$\bigcup X$} denote the union of all sets in $X$; that is, $x \in \bigcup X$ iff\ $x \in S$ for some $S \in X$. 
  
  Let $K = (V, E)$ be a finite graph, not necessarily connected.
  Two vertex subsets $L,M \subseteq V$ are \emph{connected} if at least one vertex in $L$ is adjacent to some vertex in $M$; otherwise they are \emph{independent}. %
    A family~$H = \{H_1, \dots H_s \}$ of subsets of~$V$
    is \emph{independent} if each two members from the family are independent.
\end{definition}

{
}

\fi

\begin{definition}[generalized cutset; in short, \gencutset]\label{def:generalized-cutset-final}   For $G = (V, E)$ a finite connected graph, let
\begin{compactitem}[--]

    \item $\mathcal{C}  = \{C_1, \dots, C_t\}$ be a \newH{family} of $t$ pairwise disjoint, 
    nonempty subsets of $V$, \newH{each inducing a connected subgraph},
    \item $\tau  = (\tau_1, \dots, \tau_t)$ be a sequence of natural numbers, with $\tau_j$ called $C_j$'s \emph{pass-through number} and the sum $\Sigma \tau$ of all $\tau_j$ called the \emph{valence},

    \item $G \setminus \mathcal{C}$ be the subgraph of $G$ induced by~$V \setminus \bigcup \mathcal{C}$, and %
    \item $H  = \{H_1, \dots, H_{\Sigma \tau+r}\}$ be an independent partition of $V \setminus \bigcup \mathcal{C}$ with $r\ge 2$. %
\end{compactitem}
Assume, in addition, that
\begin{compactitem}[--]
    \item for each $C_i$ and $H_j$ there is at most one vertex $s_{i,j}$ in $C_i$ adjacent to any vertices in $H_j$, with $s_{i,j}$ referred to as the \emph{contact vertex} for $C_i$ and $H_j$, 
    \item each  $C_j \in \mathcal{C}$ is either a ``type-I member'' containing one vertex, or a ``type-II member'' containing more than one, 
    \item $\tau_i = 1$ for each type-I $C_i$,%
    \item the type-II members of $\mathcal{C}$ form an independent family, and 
    \item for each type-II member $C_i$, there are $2\tau_i + 1$
    sets $H_j$ for which there exists a contact vertex  $s_{i,j} \in C_i$, 
    and these contact vertices are distinct.
\end{compactitem}
\noindent Then $\mathcal{C}$ is a \gencutset\ of gap $r \ge 2$ and valence $\Sigma \tau$, with witness~$H$. Such a cutset is \emph{tame} if it contains at most one type-II member.  
\end{definition}

Note that we \emph{do not} require any $H_j$ to be connected.

\if Our final version of the definition now allows for cutset members with pass-through numbers greater than one. \fi

\begin{observation} \label{generalize}
    Every \cutset~$\mathcal{C}$ of gap $\ge 2$ and valence $t$ satisfies \cref{def:generalized-cutset-final} %
    for \gencutset of gap $\ge 2$ with the same valence $t$, once each vertex~$c_i \in \mathcal{C}$ is converted into the corresponding singleton set $\{ c_i\}$, 
    $\tau$ is set equal to $(1,1, \dots , 1)$ with $\Sigma \tau = t$, 
    and $H$ is set equal to the set of connected components in $G\setminus \mathcal{C}$. %
\end{observation}

\noindent \cref{generalize} explains why we present only one version of the main theorem that follows.

\newcommand{\claimdeprived}{%
  For every allocation, a deprived agent exists.
}
\newcommand{\claimpriveleged}{%
  For every allocation, a privileged agent exists.
}
\newcommand{\claimdeprivedprivileged}{%
  For every allocation, a deprived agent and a privileged exist.
}
\newcommand{\claimpd}{%
  For every connected allocation, privileged agents are never deprived.
}
\begin{theorem}[Main Theorem]\label{thm:main}
  Let $G = (V,E)$ be a finite, connected graph.
  Suppose $\mathcal{C} = \{C_1, C_2, ... , C_t\}$ is a \gencutset\ for $G$, of gap $r \ge 2$ and valence $\Sigma \tau$, with witness $H = \{H_1, H_2, \dots,\\ H_{\Sigma \tau+r}\}$. Then for each integer~$n$ lying within the ``critical interval'' $\Sigma \tau < n < r+\Sigma \tau $, 
    \begin{compactitem}[--]
    \item there exist common monotone valuations for  $n$ agents, under which no connected \emph{\efone} allocations exist (whence no connected \emph{\efouter} allocations exist);
        \item if $\mathcal{C}$ is tame, there exist common additive valuations for $n$ agents, under which no connected \emph{\efone} allocations exist (whence no connected \emph{\efouter} allocations exist).
     \end{compactitem}
\end{theorem}

\begin{proof}
{We begin by introducing some necessary definitions.}
We say that a set $S \subseteq V$ \emph{dominates} a type-I member $C_i = \{c_i\} \in \mathcal{C}$ if $c_i \in S$; $S$ \emph{dominates} a type-II member $C_i \in \mathcal{C}$ if $S$ includes two or more of $C_i$'s contact vertices~$s_{i,j}$. Given an 
allocation $A =  (A_1, A_2, ... , A_n)$ of $G$'s vertices to the $n$ agents, we say that agent~$j$ dominates a member~$C_i \in \mathcal{C}$ if their assigned piece~$A_j$ dominates~$C_i$. An agent who dominates no member of $\mathcal{C}$ is \emph{deprived}. 

Choose one distinguished vertex $h_j \in H_j$ from each $H_j \in H$, so that the number of distinguished vertices is $|H|$. An agent whose share includes two or more of these distinguished vertices is said to be \emph{privileged}. We first show that deprived and privileged agents both exist, then construct valuations under which the former envies the latter by more than one item:

\begin{claim}%
  \label{claim:deprive-privileged-exists}
  \claimdeprivedprivileged
\end{claim}

\appendixproofwithstatement{claim:deprive-privileged-exists}{\claimdeprivedprivileged}
{\begin{proof}[Proof of \cref{claim:deprive-privileged-exists}]
    \renewcommand{\qedsymbol}{$\diamond$}
    Let $A=(A_1,\dots,A_n)$ be an arbitrary allocation for $n$ agents.
    Then, at most $\tau_i$ agents can dominate any single member $C_i \in \mathcal{C}$, so the number of agents who dominate members of $\mathcal{C}$ is at most $\Sigma \tau$. But there are $n > \Sigma \tau$ agents.
    This means that there must be one agent that \emph{does not} dominate any cutset member, showing the first part of the statement. 
    
    For the second part, recall that the number of agents is $n < r+ \Sigma  \tau  = |H|$.  With more distinguished vertices than agents, some agent's share $A_k$ must include two or more of the $h_j$, meaning that some agent is privileged, as desired.
  \end{proof}
}

\begin{claim}%
  \label{claim:priveleged-not-deprived}
  \claimpd
\end{claim}

   \appendixproofwithstatement{claim:priveleged-not-deprived}{\claimpd}
   {\begin{proof}[Proof of \cref{claim:priveleged-not-deprived}]
     \renewcommand{\qedsymbol}{$\diamond$}
     Let $A=(A_1,\dots,A_n)$ be an arbitrary connected allocation for $n$ agents.
     By \cref{claim:deprive-privileged-exists}, let $i$ be a privileged agent, so that $A_i$ is a connected share that contains distinguished vertices from two different members of $H$.  Let $\rho = x_1, x_2, \dots , x_{k-1}, x_k$ be a shortest path possible consisting entirely of vertices from $A_i$ and joining members $x_1$ and $x_k$ of different sets in $H$.  As $H$ is an independent collection, $x_1$ and $x_k$ are not adjacent, so there must be at least one vertex in the ``middle'' part $x_2, \dots x_{k-1}$ of $\rho$. None of those middle vertices are in $\bigcup H$, else we would get a shorter path of the desired kind, so they all come from $\bigcup \mathcal{C}$.  If any middle vertex is some $c_j \in \{c_j\} = C_j \in \mathcal{C}$ then $i$ dominates that type-I member $C_j$, so $i$ is not deprived, as desired.  If not, then all of the middle vertices come from type-II members of $\mathcal{C}$.  But the type-II members form an independent collection, so all of the middle vertices come from the same type-II member $C_j$.  Thus $x_2$ and $x_{k-1}$ are each contact vertices from the same $C_j$, but for distinct sets from $H$. We cannot have $x_2 = x_{k-1}$, because contact vertices for different sets in $H$ and the same $C_j$ are required to be distinct.  So $x_2$ and $x_{k-1}$ are two contact points from the same type-II member $C_j$, showing that \newH{$A_i$} dominates $C_j$, as desired.   
   \end{proof}
   }

   \newH{Now, we are ready to state concrete valuations for which no connected \efone\ allocations exist.} 
   We define the common valuation $v$ of a set of vertices as the sum $v = v_H + v_\mathcal{C}$ of an \emph{$H$-part} and a \emph{$\mathcal{C}$-part}.  For the $H$-part, set $v_H (h_j) = 1$ for each distinguished vertex $h_j$, and  $v_H (y) = 0$ for every other vertex $y \in \bigcup H$.   Then $v_H (A_i)$ is the sum of these values for all vertices in the set $A_i \cap (\bigcup H)$.  The definition of $v_\mathcal{C}$ depends on whether $\mathcal{C}$ is tame:

    \begin{compactitem}[-]
      \item[Case 1: \emph{$\mathcal{C}$ is tame.}]  Then for each $x \in \bigcup \mathcal{C}$ we set $v_\mathcal{C} (x) = 1$ if $\{x \}$ is a type-I member of $\mathcal{C}$, $v_C (x) = \frac{1}{3}$ if $x$ is one of the contact vertices in the only type-II member of $\mathcal{C}$, and $v_\mathcal{C} (y) = 0$ for every other vertex $y \in \bigcup \mathcal{C}$. Then $v_\mathcal{C} (A_i)$ is defined to be sum of these values for all vertices in the set $A_i \cap (\bigcup \mathcal{C})$.  In this case, $v = v_H + v_\mathcal{C}$ is additive as well as common and the total value of any deprived agent's share $A_j$ is at most $1 + \frac{1}{3}$; see \cref{claim:priveleged-not-deprived}.  
      
      \item[Case 2: \emph{$\mathcal{C}$ is not tame.}]  Then $v_\mathcal{C} (A_i)$ is defined to be the number of members of  $\mathcal{C}$ dominated by $A_i$. In this case, $v = v_H + v_\mathcal{C}$ is still CM, but is not additive (because one or more contact vertices, each from a different type-II member, add no value to a share, whereas two or more contact points from the same type-II member adds $1$ to the value). As $v_C$ contributes no value to the share of a deprived agent, and deprived agents are not privileged, the total value of any deprived agent's share $A_j$ is at most $1$.  
    \end{compactitem}
    
    \noindent Let $A=(A_1,\dots, A_n)$ be a connected allocation.
    We know that some privileged agent $k$ receives a share~$A_k$ containing two or more distinguished vertices, each of value $1$. By \cref{claim:priveleged-not-deprived} we know that $k$ is not deprived, so $A_k$ dominates some member of the cutset, by including either some one vertex $x$ with $\{x \}$ being a type-I member of $C$, or some two contact vertices from the same type-II member $C_j$ of $\mathcal{C}$. 
    
    In Case 1 we conclude that $A_k$ has (additive) value of at least $1 + 1 + \frac{1}{3} + \frac{1}{3}$  or $1 + 1 + 1$, so removing any single vertex leaves $A_k$ with value at least $1\frac{2}{3}$.  The deprived agent thus envies agent $k$ by more than $1$ item, as his own share is worth at most $1\frac{1}{3}$ for Case 1.
    
    In Case 2 we conclude that $v_H$ awards a value of at least  $1 + 1$ to $A_k$ with $v_\mathcal{C}$ providing an additional value of at least~$1$, for a total of at least~$3$. Removing any single vertex would not reduce that value below $2$.  The deprived agent again envies agent $k$ by more than $1$ item, as his own share is worth at most $1$ in Case 2.
\end{proof}

Note that requiring $r \ge 2$ in \cref{thm:main} guarantees that the critical interval contains at least one integer $n$.
\newH{Further, two restrictions in the \gencutset definition---that cutset members induce connected subgraphs and that pass-through numbers $2\tau_j + 1$ be odd---are not used in the proof of \cref{thm:main}, but help to reduce the search space for cutsets. As we see next, the imposition of these two restrictions does not eliminate any cases to which the theorem applies.

\todoBinline{Note two changes/clarifications in wording (above). First is ``---that cutset members induce connected subgraphs and that pass-through numbers $2\tau_j + 1$ be odd---are not used in the proof of \cref{thm:main}.'' The previous wording allowed for the possible misreading that one might need an alternative version of the proof of \cref{thm:main} to show that it still goes through without these restrictions. This should clarify why  no proof is needed to show that the Theorem holds without the restrictions. I also think we need no formal statement saying this is the case Second is "we state and prove a proposition," allowing us to have the formal statement of the \emph{other} proposition---saying the restrictions lead to no loss of cases to which the main theorem applies---appear only in Appendix (see \ref{appenxi:cust-relax}).}

}

\toappendix{
  \subsection{Two optional restrictions in the cutset definition}\label{appenxi:cust-relax}
The phrase ``relaxed version of Definition \ref{def:generalized-cutset-final},'' as used below, refers to the following modification of the definition for graph cutset:
\begin{quote}
  A type-II member $C_i$ of $\mathcal{C}$ is \emph{no} longer required
\begin{compactenum}[(R1)]
  \item\label{relaxed1} to induce a connected subgraph, or
  \item\label{relaxed2} to contain an odd number $2\tau_i + 1$ of contact vertices $s_{i,j} $---an even number $2\tau_i$ is also permitted.
\end{compactenum}
\end{quote}
We show here that the additional ``relaxed'' cutsets admitted by these loosened requirements yield no consequences for connected \eefone fair division beyond those that already follow from the narrower class of ``restricted'' cutsets that meet the original definition. In particular, the condition $I_{\mathcal{C}} \subseteq I_{\mathcal{C}'}$ in Proposition \ref{OddandConnected} (below) implies, for any connected graph $G$ and integer $n$, that if some relaxed cutset $\mathcal{C}$ can be applied, via \cref{thm:main}, to show that connected EF1 allocations are not guaranteed for $n$ agents, then some restricted cutset $\mathcal{C}'$ can similarly be applied to reach the same conclusion.

In what follows, it is helpful to recall the notation of~$\bigcup X$. Thus, while $|\mathcal{C}|$ denotes the number of members of the cutset $\mathcal{C}$, $\bigcup \mathcal{C}$ stands for the set of vertices contained in the union of those members; $|H|$ and $\bigcup H$ have analogous meanings for $H$ ($\mathcal{C}$'s witness partition.)

\begin{proposition} \label{OddandConnected} Let $G = (V,E)$ be a connected graph and $\mathcal{C}$ a \gencutset, with \newH{pass-through numbers~$\tau$} and witness partition~$H$, satisfying the relaxed version of Definition \ref{def:generalized-cutset-final} (see (R\ref{relaxed1})--(R\ref{relaxed2})) for $G$.  Let $I_{\mathcal{C}}$%
  denote the critical interval of integers $n$ with $\Sigma \tau < n < |H| = \Sigma \tau + r$, \newH{i.e., $I_{\mathcal{C}} = \{\Sigma \tau+1, \dots, \Sigma\tau+r-1\}$.}
Then there exists a 
cutset $\mathcal{C}'$ with valence $\Sigma \tau '$, witness partition $H'$, and critical interval $I_{\mathcal{C}'}$ for which:
\begin{compactenum}[(1)]
\item\label{prop-valence} $\Sigma \tau ' = \Sigma \tau $ and $|H'| \ge |H|$, so that
 $I_{\mathcal{C}} \subseteq I_{\mathcal{C}'}$, and
\item\label{prop-original} $\mathcal{C}'$ satisfies the original version of Definition \ref{def:generalized-cutset-final}.
\end{compactenum}
\end{proposition}

\begin{proof}
Given a graph~$G$ and a \gencutset~$\mathcal{C}$ as described above, assume that some type-II member $C_i \in \mathcal{C}$ fails to satisfy at least one of the two restrictions under discussion.  It is enough to construct a \gencutset~$\mathcal{C}^\star$ that continues to meet the relaxed version of the definition, and additionally satisfies:

\begin{compactenum}[(1$^*$)]
\item\label{mod-valence} $\Sigma \tau ^\star = \Sigma \tau $ and $|H^\star| \ge |H|$, so that
 $I_{\mathcal{C}} \subseteq I_{\mathcal{C}^\star}$, 
\item\label{mod-cutsetmember} $|\mathcal{C}^\star | \ge |\mathcal{C}|$ and \newH{$\bigcup H^\star \supseteq \bigcup H $},  and either\\ $|\mathcal{C}^\star | > |\mathcal{C}|$\hspace{1mm} or \hspace{0.4mm} \newH{$\bigcup H^\star \supsetneq \bigcup H $}.
\end{compactenum}

\noindent  Here condition (\ref{mod-valence}$^*$) mirrors condition \eqref{prop-valence}.
Condition (\ref{mod-cutsetmember}$^*$) asserts that the $^\star$-operation never decreases the number of members of the cutset, and never \newH{remove vertices from the union of all subsets in the witness partition; moreover, this operation either strictly increases the number of members in the cutset or strictly enlarges the set of the vertices in the witness partition.} %
(In practice, we will see that these changes happen when a member of the cutset is split in two, or when vertices are transferred from $\bigcup \mathcal{C}$ to $\bigcup H$, with no cutset member ever disappearing completely, and no vertex transfers ever going in the opposite direction.)

This $\mathcal{C}^\star$ either satisfies the original definition, or else we can iterate the construction, generating a chain  $\mathcal{C}^\star$, $ \mathcal{C}^{\star \star}$,  $ \mathcal{C}^{\star \star \star}$, \dots. This chain only terminates when the terminal cutset $ \mathcal{C}^{\star \star \dots \star}$ satisfies the original, restricted definition. Moreover, it must terminate because neither the number of cutset members, nor the number of vertices contained by all sets in the witness partition, can grow beyond the number of vertices of the finite graph $G$. The construction of $\mathcal{C}^\star$ from $\mathcal{C}$ now proceeds via cases.

\begin{compactenum}
  \item[Case 1:] Assume that $C_i$ has an even number $2\tau_i \ge 4$ of contact vertices, and let $s_{i,j}$ be one of them.  Construct $\mathcal{C}^\star$ from $\mathcal{C}$ as follows: remove $C_i$ from $\mathcal{C}$, and replace it with two new members: $C_i \setminus \{ s_{i,j}\}$ (which is type-II, with $2\tau_i^\star + 1 = 2(\tau_i - 1) +1$ contact vertices) and  $\{ s_{i,j}\}$ (which is type-I); note that the original $C_i$ makes a  contribution to total valence equal to the combined contributions of its two replacements $C_i \setminus \{ s_{i,j}\}$ and $\{ s_{i,j}\}$. Set $H^\star = H$.  Then $\mathcal{C}^\star$ satisfies the relaxed definition of cutset, with $\Sigma \tau ^\star = \Sigma \tau $, \newH{$H^\star = H$}, and $|\mathcal{C}^\star| = |\mathcal{C}| + 1$. %

  \item[Case 2:] Assume that $C_i$ has $2$ contact vertices, $s_{i,j}$ and $s_{i,k}$ (with $j \neq k$).  Construct $\mathcal{C}^\star$ from $\mathcal{C}$ as follows: remove $C_i$ from $\mathcal{C}$, and replace it with the new type-I member $ \{ s_{i,j}\}$. Note that the original $C_i$ and its replacement make the same contribution of $1$ to the valence. To construct $H^\star$ from $H$, move all vertices in $C_i \setminus \{ s_{i,j}\}$ into $H_k$.   Then $\mathcal{C}^\star$ satisfies the relaxed definition of cutset,
  with $\Sigma \tau ^\star = \Sigma \tau $, $| H^\star| = | H|$, $|\mathcal{C}^\star| = |\mathcal{C}|$, and \newH{$\bigcup H^\star \supsetneq \bigcup H$}.

  \item[Case 3:] Assume that $C_i$ has an odd number $2\tau_i +1 \ge 3$ of contact vertices, induces a subgraph of $G$ having more than one connected component, and one of these connected components $D$ contains none of $C_i$'s contact vertices.   To construct $\mathcal{C}^\star$ from $\mathcal{C}$, remove $C_i$ from $\mathcal{C}$, and replace it with $C_i \setminus D$, a type-II member containing the same number $2\tau_i +1$ of contact vertices.  To construct $H^\star$ from $H$, add $D$ to $H$ as a new piece of the partition. Then, $\mathcal{C}^\star$ satisfies the relaxed definition of cutset,
  with $\Sigma \tau ^\star = \Sigma \tau $, $| H^\star| = | H|+1$, $|\mathcal{C}^\star| = |\mathcal{C}|$, and \newH{$\bigcup H^\star \supsetneq \bigcup H$}.

\item[Case 4:] Assume that $C_i$ has an odd number $2\tau_i +1 \ge 3$ of contact vertices, induces a subgraph of $G$ having more than one component, and does not fit Case 3, but some connected component $D$ of $C_i$  contains exactly one contact vertex $s_{i,j}$ of $C_i$.   To construct $\mathcal{C}^\star$ from $\mathcal{C}$, remove $C_i$ from $\mathcal{C}$, and replace it with $C_i \setminus D$, a type-II member containing $2\tau_i $ contact vertices, so that the original~$C_i$ and its replacement both make the same contribution of $\tau_i$ to total valence. To construct $H^\star$ from $H$, add all vertices of $D$ into $H_j$.
Then, $\mathcal{C}^\star$ satisfies the relaxed definition of cutset,
with $\Sigma \tau ^\star = \Sigma \tau $, $| H^\star| = | H|$, $|\mathcal{C}^\star| = |\mathcal{C}|$, and \newH{$\bigcup H^\star \supsetneq \bigcup H$}. 

\item[Case 5:] Assume that $C_i$ has an odd number $2\tau_i +1 \ge 3$ of contact vertices, induces a subgraph of $G$ having more than one component, and each of these components contains at least two contact vertices of $C_i$. Let $D$ be any component having an odd number $2\tau_i^\dagger+1 \ge 3$ of contact vertices, so that $C_i \setminus D$ contains $2\tau_i^{\dagger \dagger}$ contact points with $2\tau_i+1 = \big(2(\tau_i^{\dagger})+1\big) + \big(2\tau_i^{\dagger\dagger}\big)$. To construct $\mathcal{C}^\star$ from~$\mathcal{C}$, remove $C_i$ from $\mathcal{C}$ and replace it with the two type-II members $D$ and $C_i \setminus D$, noting that the original~$C_i$ makes a  contribution $\tau_i$ to total valence equal to the combined contributions $\tau_i^{\dagger} + \tau_i^{\dagger\dagger}$ of its two replacements $D$ and $C_i \setminus D$.  Set $H^\star$ = $H$.
Then, $\mathcal{C}^\star$ satisfies the relaxed definition of cutset,
with $\Sigma \tau ^\star = \Sigma \tau $, \newH{$H^\star  = H$}, and $|\mathcal{C}^\star| = |\mathcal{C}| + 1$. %
\qedhere\end{compactenum}
\end{proof}
}

\subsection{A Counterexample}

Consider the following three conditions on a finite and connected graph $G$:
\begin{compactenum}[(C1)]
    \item\label{traceable} $G$ is traceable.
    \item\label{EF1} $G$ guarantees connected \emph{\efouter} allocations \emph{universally}, for CM valuations.
    \item\label{cutset} $G$ contains no generalized cutsets of gap $\geq 2$.
\end{compactenum}

\noindent We know the following implications: ``(C\ref{traceable}) $\Rightarrow$~(C\ref{EF1})'' (from \cref{IgThm}) and ``(C\ref{EF1}) $\Rightarrow$ (C\ref{cutset})'' (from \cref{thm:main}).  At one point, we did not know whether either arrow reversed. Recently we found a $10$-vertex graph showing that ``(C\ref{cutset}) $\Rightarrow$ (C\ref{traceable})'' fails.
This was not a great surprise as ``(C\ref{cutset}) $\Rightarrow$ (C\ref{traceable})'' would imply that ``(C\ref{cutset}) $\Leftrightarrow$ (C\ref{traceable})'', and hence NP is contained in coNP; the latter, however, is widely considered as unlikely. The reason is that checking traceability is an NP-complete problem, while checking the non-existence of generalized cutsets is a coNP problem; see next section for the complexity of the latter problem.

The example was a bit too large to check directly whether connected \emph{\efouter} allocations were guaranteed universally.  Much more recently we were able to reduce the earlier example to the $8$-vertex graph \emph{JCS} presented here, which was small enough to yield to a trial-and-error search for a ``bad'' common additive valuation. It is worth noting that while the valuation used in the proof for \emph{JCS} is not very complicated, it does seem quite different from the valuations used (in the previous section) to defeat \emph{\efouter} allocations in graphs that contain cutsets.

\begin{theorem} \label{JCScounterexample} The non-traceable \emph{JCS} graph of \cref{fig:VII+JCS+Lips-graph} has no \gencutset{s} of gap $\geq 2$ (of any kind), yet fails to guarantee connected \emph{\efouter} allocations for three agents, even for agents with CA valuations. Thus ``(C\ref{cutset}) $\Rightarrow$ (C\ref{EF1})'' fails.    
\end{theorem} 

\newcommand{\claimthreepartition}{%
  Let $P$ be a $3$-partition, $A$ denote the piece of $P$ containing vertex $a$ in \cref{fig:VII+JCS+Lips-graph}, and $H$ denote the piece containing vertex $h$. Then for $P$ to avoid a failure of type $X >> Y$ we must have that $A \neq H$, with
  \begin{compactenum}[(i)]
    \item $A$ containing $b$ and $c$, and omitting $d$ and $e$, and
    
    \item $H$ containing $g$ and $f$, and omitting $d$ and $e$.
  \end{compactenum}
}

\begin{proof} 
One can check by inspection that the JCS graph contains no cutsets of gap $\geq 2$, and it is also easy to see that no Hamiltonian path exists (which, alternately, follows from \cref{IgThm}).  To see that connected \emph{\efone} allocations may fail to exist, consider the vertex weights appearing directly above the vertices in \cref{fig:VII+JCS+Lips-graph}. Let the common value $v(S)$ assigned (by all agents) to a connected set $S$ of vertices be given by the sum of the weights of the vertices in $S$.  

A partition $P$ of the JCS vertex set V into three connected pieces will be called a $3$-\emph{partition.}  Let $X$ and $Y$ be  pieces of a $3$-partition $P$, and $X^\star$ denote the set of vertices that remain in $X$ after $X$'s most valuable vertex is removed.  We will write $X >> Y$ if $v(X^\star) > v(Y)$. If a $3$-partition $P$ contains two such pieces $X$ and $Y$, then any assignment of $P$'s pieces to the agents clearly fails to be  \emph{\efone} (whence \emph{\efouter} also fails).  In this case we will say simply that $P$ \emph{fails}. Our goal, then, will be to show that every $3$-partition fails in this way. 

\begin{claim}%
  \label{clm:3partition}
\claimthreepartition
\end{claim}

\appendixproofwithstatement{clm:3partition}{\claimthreepartition}{
  \begin{proof}[Proof of \cref{clm:3partition}]
    \renewcommand{\qedsymbol}{$\diamond$}
    To prove the claim, note that if $A = \{a\}$ or $A = \{a,b\}$ then $v(A)$ is only $2$ or $4$, and the value of the remaining vertices in $V \setminus A$ is at least $14$.  But there exists no partition of $V \setminus A$ into two connected pieces of value $7$ each, so one of these pieces $X$ must have value at least $8$, whence $X >> A$.  So $A$ contains $a$, $b$, and at least one more vertex.  Similarly, $H$ contains $h,g$  and at least one more vertex.  If $A$ contains $g$ then $A = H$, with $v(A) \geq 8$.  But then the value of the remaining vertices is only $10$ -- small enough to force $A >> Y$ for some $Y \in P$.  So for $P$ to avoid failure we must have $a, b,$ and $c$ in $A$; $f, g,$ and $h$ in $H$; and $A \neq H$.  If $A$ also contained $d$ or $e$, we would again get $A >> Y$ for some $Y \in P$, and the same reasoning applies to $H$.  This establishes the claim.
  \end{proof}
}

\noindent The theorem follows from \cref{clm:3partition} because the third piece of P is forced to be $\{ d, e\}$, which is disconnected.  \end{proof}

\section{NP-Hardness of Finding Graph Cutsets}\label{sec:cutset:NP-hard}
\appendixsection{sec:cutset:NP-hard}
We begin with a reduction showing that finding an elementary \cutset is computationally hard, then modify it for our more general \gencutset notion.

\newcommand{\cutsetnph}{%
  It is NP-complete to decide whether an undirected graph admits an\\ \cutset{} of valence~$t$ and gap~$\ge 2$.
}
\begin{theorem}%
  \label{NP-hard:cutset}
  \cutsetnph
\end{theorem}

\begin{proof}%
  NP-containment is straightforward since one can check in linear time whether a given subset of $t$ vertices is an elementary cutset of valence~$t$ and gap~$\ge 2$.
  To show NP-hardness, we reduce from the NP-complete \textsc{Clique} problem~\cite{GJ79}.
  This problem has as input an undirected graph~$\hat{G}=(\hat{V}, \hat{E})$ and a number~$h$.
  The task is to decide whether there exists a vertex subset~$V'\subseteq \hat{V}$ of size~$h$ which induces a complete subgraph (aka.\ \emph{clique}), i.e., one for which each two vertices in $V'$ are adjacent. 

  Let $I=(\hat{G}=(\hat{V}, \hat{E}), h)$ denote an instance of \textsc{Clique} with $\hat{V}=\{v_1,\ldots,v_{\hat{n}}\}$ and $\hat{E}=\{e_1,\ldots, e_{\hat{m}}\}$.  
  We construct a graph~$G$ for the cutset problem as follows.
  \begin{compactitem}[--]
    \item For each edge~$e\in \hat{E}$, create an \emph{e-vertex} called~$u_e$.
    \item For each vertex~$v_i\in \hat{V}$, create a \emph{v-vertex} called~$v_i$.
    All v-vertices form a huge clique while each e-vertex~$u_e$ is only adjacent to the v-vertices that correspond to the endpoints of~$e$.
    \item Finally, create $\binom{h}{2}+h+1$ dummy vertices~$a_1,\ldots, a_{\binom{h}{2}}$ and $b_1,\dots, b_{h+1}$ such that each vertex~$b_{z'}$ is only adjacent to each vertex~$a_z$, while each~$a_z$ is adjacent to all other vertices. Let $A=\{a_z\mid z\in [\binom{h}{2}]\}$ and $B=\{b_1,\dots,b_{h+1}\}$.
  \end{compactitem}
  To complete the construction, let $t=\binom{h}{2}+h$.
\appendixcorrectnessproofwithstatement{The correctness proof can be found in \cref{proof:NP-hard:cutset}}{NP-hard:cutset}{\cutsetnph}{
  We continue to show that the construction %
  is correct, i.e.,
  graph~$\hat{G}$ has a clique of size~$h$ if and only if $G$ has cutset  of valence~$t$ deleting which yields at least $t+2$ connected components.
  
  For the ``only if'' direction, it is straightforward to verify that deleting all vertices in $A$ together with the $h$ v-vertices which correspond to the size-$h$ clique results in $\binom{h}{2} + h+2 = t+2$ connected components.
  These connected components are $\binom{h}{2}$ isolated e-vertices, the $h$ dummy vertices from~$B$, and the rest.
  
  For the ``if'' direction, let $U'\subseteq V(G)$ denote a subset of $t$ vertices deleting which yields at least~$t+2=\binom{h}{2}+h+2$ connected components.
  Let $G'$ denote the resulting graph~$G-U'$.
  First, observe that every dummy~$a_z$ must be in the cutset~$U'$ since every vertex is adjacent to~$a_z$.
  This means that $A\subseteq U'$, and each $b_{z'}$ is a single connected component in~$G'$.
  There remain $h$ vertices in~$U'\setminus A$ and $\binom{h}{2}+1$ connected components in $G'-B$ to be determined.
  Since there are more than $h$ v-vertices and all v-vertices are adjacent among each other, we cannot delete $h$ vertices to disconnect any two v-vertices.
  This implies that among the $\binom{h}{2}+1$ connected components in $G'-B$, there is exactly one which intersects the v-vertices.
  Let $X$ denote this component. 
  If two e-vertices are in the same component, then they are in~$X$ since they are only adjacent to their endpoint v-vertices and each component in~$G'$ is connected. 
  In other words, there must be at least $\binom{h}{2}$ isolated e-vertices left in~$G'-B$.
  We claim that these $\binom{h}{2}$ isolated e-vertices correspond to a clique in the original graph~$\hat{G}$.
  To show this, it suffices to show that there are $h$ endpoints of the edges corresponding to the $\binom{h}{2}$ e-vertices.
  To this end, let $V'_E=\{u_e \mid u_e \text{ is an isolated vertex in } G'\}$.
  Since $|V'_E|\ge \binom{h}{2}$, the corresponding edges have in total at least $h$ end points such that the lower bound is reached only if they correspond to a size-$h$ clique.
  By previous reasoning that there remaining $h$ vertices in $U'\setminus A$, we can delete at most $h$ many v-vertices.
  Hence, these deleted v-vertices must correspond to a size-$h$ clique.%
}%
\end{proof}

The reduction behind \cref{NP-hard:cutset} introduces an enforcement gadget (the large clique) and uses a counting argument that only works for the case when all cutset members are type-I (see \cref{def:generalized-cutset-final}).
This argument alone does not work for the more general cutset definition with type-II members, however, since a type-II cutset member can use a valence that accounts for half the number of ``disconnected components'' in the remaining graph. %
To amend this, the argument uses a second type of enforcement gadget to show:

\newcommand{\gencutsetnph}{%
  It is NP-complete to decide whether an undirected graph admits a \gencutset
  of valence~$t$ and gap~$\ge 2$.
}
\begin{theorem}%
  \label{thm:gencutset-NPh}
  \gencutsetnph
\end{theorem}

\appendixproofwithstatement{thm:gencutset-NPh}{\gencutsetnph}{
\begin{proof}
  We adapt the construction given in the proof of \cref{NP-hard:cutset}.
  So we will again reduce from \textsc{Clique}.

  Let $I=(\hat{G}=(\hat{V}, \hat{E}), h)$ denote an instance of \textsc{Clique} with $\hat{V}=\{v_1,\ldots,v_{\hat{n}}\}$ and $\hat{E}=\{e_1,\ldots, e_{\hat{m}}\}$.  
  Without loss of generality, assume that $h$ is odd. 
  We construct a graph~$G$ for the cutset problem as follows.
  \begin{compactitem}[--]
    \item For each edge~$e\in \hat{E}$, create an \emph{e-vertex} called~$u_e$.
    \item For each vertex~$v_i\in \hat{V}$, create a \emph{v-vertex} called~$v_i$ and a copy of it, called~$w_i$.
    Every $w_i$ is only adjacent to its primary~$v_i$.
    All v-vertices form a huge clique, while each e-vertex~$u_e$ is only adjacent to the v-vertices that correspond to the endpoints of~$e$ and to the a-vertices that we specify next.
    For the sake of brevity, let $W=\{w_i\mid i \in [\hat{n}]\}$.
    \item Finally, create two groups of dummy vertices, where $L$ is an integer larger than~$\hat{m}$.
    \begin{itemize}[$\bullet$]
      \item Create $L+\hat{n}+\binom{h}{2}+2$ dummy vertices, called~$a_1,\ldots, a_{L+\hat{n}+\binom{h}{2}}$ and $b_1, b_{2}$ such that each vertex~$b_{z'}$ is only adjacent to all dummy vertices~$a_z$, while each~$a_z$ is adjacent to all other vertices.
      Define $A=\{a_z\mid z\in [L+\hat{n}+\binom{h}{2}]\}$ and $B=\{b_1,b_2\}$.
      \item Create $2(2L+\hat{n}+h)$ dummy vertices, called~$x_1,\dots,x_{2L+\hat{n}+h}$ and $d_1,\ldots, d_{2L+\hat{n}+h}$.
      Define $X=\{x_z\mid z\in [2L+\hat{n}+h]\}$ and $D=\{d_z\mid z\in [2L+\hat{n}+h]\}$.
      Each dummy~$d_z$ is only adjacent to the corresponding dummy~$x_z$,
      while all dummies~$x_z$ are adjacent to each other and to all
      v-vertices. 
    \end{itemize}
  \end{compactitem}
  Let the valence be~$\Sigma\tau=2L+2\hat{n}+\binom{h}{2}+h$.

  It remains to show that graph~$\hat{G}$ has a clique of size~$h$ if and only if $G$ has a \gencutset of valence~$\Sigma\tau$ and gap at least~$2$.

  For the ``only if'' direction, let $U'$ denote a size-$h$ vertex subset which induces a clique with $U'=\{v_{j_1}, \dots, v_{j_h}\}$.
  Let $R$ denote the set consisting of the v-vertices and e-vertices not from the clique.
  It is straightforward to verify that the following collection~$\{\{a_1\},\dots, \{a_{L+\hat{n}+\binom{h}{2}}\}, X\cup R, \{v_{j_1}\}, \dots, \{v_{j_h}\}\}$ is a cutset of valence $\tau$, deleting which yields $\tau+2$ disconnected subgraphs:
  the $2L+\hat{n}+h$ dummies from $D$, the $2$ dummies from $B$, the $\hat{n}$ copies of the v-vertices, and the $\binom{h}{2}$ e-vertices corresponding to the clique.

  For the ``if'' direction, let $\mathcal{C}$ be a \gencutset\ of valence~$\Sigma\tau$ and gap at least~$2$, and $H={H_1,\ldots,H_{\Sigma\tau+2}}$ the corresponding witness.
    Our method for showing that $\hat{G}$ 
    admits a size-$h$ clique is to demonstrate that any such cutset $\mathcal{C}$ and witness $H$ must necessarily resemble the ones we just described in the ``if direction''. In particular, it will turn out that~$\bigcup H$ must consist} 
    of $\binom{h}{2}$ e-vertices that correspond to the edges of a size-$h$ clique,
    the $2L+\hat{n}+h+2$ dummies from $D\cup B$, and all $\hat{n}$ vertices~$w_i$, $i\in [\hat{n}]$.
    Note that since $H$ must contain at least $2L+2\hat{n}+\binom{h}{2}+h+2$ members, it will follow that each member in $H$ must be a singleton.
    As for the \gencutset~$\mathcal{C}$, we will see that $\mathcal{C}$ consists of $\Sigma\tau - 1$ type-I members and one type-II member.
    The $\Sigma\tau - 1$ type-I members are $\{a_z\}$, $z\in [L+\hat{n}+{\binom{h}{2}}]$ and $h$ v-vertices~$v_{i_z}$, $z\in [h]$.
    These $h$-vertices will disconnect the $\binom{h}{2}$ e-vertices from $H$ and hence must necessarily form a size-$h$ clique by our construction.
    The contact vertices in the unique type-II member will be all dummies from~$X$ together with the remaining $n-h$ v-vertices.
    In total, $\mathcal{C}$ will contain $2L+2\hat{n}+\binom{h}{2}+h$ contact vertices.
    
    To support the above description, we proceed with a sequence of claims.
  \begin{claim}\label{clm:A}
    Every dummy vertex~$a_z$ forms a type-I member in $\mathcal{C}$.
  \end{claim}

  \begin{proof}
    [Proof of \cref{clm:A}]
    \renewcommand{\qedsymbol}{$\diamond$}
    By construction, every vertex~$a_z$ is adjacent to all other vertices.
    If $a_z$ were not in {$\bigcup \mathcal{C}$}, then there can be at most one member of $H$, the witness.
    If $a_z$ were in a type-II member~$C_i$ of $\mathcal{C}$, then {as different members~$H_j$ in the witness must have distinct contact vertices in $C_i$, there again could be at most one $H$-member.}
    Hence, $a_z$ can only be in a type-I member of $C_i$, as desired.
  \end{proof} %

  \begin{claim}\label{clm:X}
    Every dummy from $X$ is in~$\bigcup \mathcal{C}$. 
  \end{claim}
  \begin{proof}
    [Proof of \cref{clm:X}]
    \renewcommand{\qedsymbol}{$\diamond$}
    Suppose not.
    Then, since every dummy vertex from $X$ is adjacent to every other vertex from~$X$, 
    {no two vertices from~$X\cup V(\hat{G})$ can be in the same cutset member $D$ of $\mathcal{C}$; if they were, then all dummies from $D$ must be also in the cutset as well since they would both be contact vertices for $D$ and whatever member of the witness partition $H$ contained the vertex from $X$ lying outside $\bigcup \mathcal{C}$, but such contact vertices must be unique.}  
    By \cref{clm:A}, {the vertices of $A$ already contribute $L + \hat{n} + \binom{h}{2}$ towards the total valence $\Sigma \tau = 2L+ 2\hat{n}+ \binom{h}{2} + h$, so cutset~$\mathcal{C}$ cannot contain more than $L+\hat{n}+ h$ vertices from $X\cup V(\hat{G})$ without making $\Sigma \tau$ too great.}
    These $L+\hat{n}+h$ vertices can ``disconnect'' at most $L+2\hat{n}+\hat{m}+h$ subsets~$H_z$: $L+\hat{n}+h$ dummies from~$D$, $\hat{m}$ e-vertices, and the copy of each v-vertex.
    This is, however, less than $\tau$ since $L> \hat{m}$, a contradiction.    
  \end{proof}

 \begin{claim}\label{clm:1typeII-v}
    $\mathcal{C}$ admits a type-II member~$C''$ such that $C''$ contains at least two vertices of $X$ and at least two v-vertices, with no type-II member other than $C'$ containing any v-vertices or vertices from $X$. Moreover, every v-vertex is in $\bigcup\mathcal{C}$.
  \end{claim}

  \begin{proof}
    [Proof of \cref{clm:1typeII-v}]
    \renewcommand{\qedsymbol}{$\diamond$}
    By \cref{clm:X}, we know that every dummy from~$X$ must be in $\bigcup\mathcal{C}$.
    Hence, again by \cref{clm:A}, at least two dummies from $X$ must be in the same cutset member (which must be type-II) as otherwise the valence would exceed $\Sigma\tau$.
    Let $C'$ denote a type-II member in $\mathcal{C}$ containing at least two dummies from $X$. 
    Consequently, every v-vertex must also be in $\bigcup\mathcal{C}$, as otherwise it would have two contact vertices in~$C'$; this shows the last part of the claim.
    Moreover, as type-II members are independent, no type-II members other than $C'$ can contain any v-vertices or any vertices of $X$.

  \end{proof}
  
  Let $h'$ be the number of type-I members which consists of a single v-vertex.
  We claim that $h'\le h$.
  By \cref{clm:1typeII-v}, let $C'$ be the unique type-II member which contains more than one v-vertex and more than one dummy from $X$.
  Define $n_x=|C'\cap X|$ and $n_v=|C'\cap V|$.
  Then, the valence induced by $X$ and $V$ is at least
  \begin{align*}
    \frac{n_x+n_v}{2}+ (2L+\hat{n}+h - n_x) + (\hat{n} - n_v) 
    = 2L+2\hat{n}+h-\frac{n_x+n_v}{2}.
  \end{align*}
  By \cref{clm:A} and the valence bound~$\Sigma\tau$, we infer that
  \begin{align*}
    &~    2L+2\hat{n}+h-\frac{n_x+n_v}{2} \le L+\hat{n}+h\\
    \Leftrightarrow &~ 2L+2n-n_v \le n_x.
  \end{align*}
  Since $n_x\le |X| = 2L+\hat{n}+h$, we further infer that $n_v \ge \hat{n}- h$.
  This implies that $h' = \hat{n}-n_v \le h$.
  Now, since $C'$ can only disconnect at most $2L+2\hat{n}+h-h'$ subsets: $2L+\hat{n}+h$ from the dummies~$D$ and {$\hat{n}-h'$ due to the contact vertices from $V(\hat{G})$}, we have that the $h'$ type-I members from $V(\hat{G})$ must disconnect at least {$\binom{h}{2}+h'$} subsets, $h'$ of which coming from the copies~$w_{i}$.
  \newH{Note that by \cref{clm:A,clm:X,clm:1typeII-v} no vertex in the union~$D\cup W\cup B \cup E(\hat{G})$ can disconnect any subsets since the neighbors of each vertex in the union are already in the cutset.}
  Hence, the remaining $\binom{h}{2}$ subsets are e-vertices, which is only possible if
  they correspond to a clique of size~$h'=h$ {since $h' \le h$}, as desired.
\end{proof}

\section{EF1 Spectrum of A Graph}\label{sec:spectrum}
\appendixsection{sec:spectrum}
Suppose that we fix a finite graph $G=(V,E)$ along with some class $\mathcal{V}$ of valuations (all monotone valuations, for example) and ask: for which natural numbers $n$ are \emph{\efouter} allocations of $G$ guaranteed for $n$ agents with valuations in~$\mathcal{V}$?  We will record the answer in the form of an infinite sequence of \emph{yes-no} answers, with the $n^{\emph th}$ member of the sequence being a ``\emph{yes}'' if the \emph{\efouter} guarantee holds for $n$, and a ``\emph{no}'' if the \emph{\efone} guarantee fails for $n$.  We will refer to that sequence as the \emph{\efouter} \emph{spectrum} of $G$ for the class $\mathcal{V}$ (or just as the \emph{spectrum}, when the context is clear). For example, we know from \cref{IgThm} that the spectrum of a traceable graph is $\langle${\emph{yes}, \emph{yes}, \dots, \emph{yes}, \dots}$\rangle$.

What general patterns hold for the spectra of all graphs? %
We begin with two positive results that make use of a \emph{picking order} when the number of agents is large enough relative to the size of the graph. The first holds for any connected graph whenever the number of agents is at least one less than the number of vertices; the second holds for graphs meeting a simple condition, when the number of agents is two less than the number of vertices. In combination with our main Theorem \ref{thm:main} these allow us to pin down the complete spectrum for a few graphs, and the common pattern they exhibit in turn suggests a general conjecture.  

\newcommand{\obsnminusone}{%
  For connected graphs with $|V|$ vertices and under monotone preferences,
  there always exists a connected \efouter\ allocation for $n \ge |V|-1$ agents.
}
\begin{observation}%
  \label{obs:|V|-1}
  \obsnminusone
\end{observation}

\appendixproofwithstatement{obs:|V|-1}{\obsnminusone}{
\begin{proof}
  If $n=|V|-1$, we order the agents arbitrarily and let each in the order pick his most preferred vertex among those still available after agents earlier in the order have picked. The one remaining vertex %
  is no more valuable to any agent than the vertex picked earlier by that agent, so it can be given to any agent for which connectivity is maintained.
  The resulting allocation is connected and \emph{\efouter} for arbitrary monotone preferences.
  
  If $n \ge |V|$ of vertices, we can give each vertex to a different agent (with some agents possibly getting no vertices); the result is \emph{\efouter} regardless of $\mathcal{V}$.
\end{proof}
}
When $n=|V|-2$, we also identify a large class of connected graphs which guarantee the existence of connected \efouter\ allocations for $n$ agents.
That class includes all %
connected graphs for which no two degree-one vertices share the same neighbor. 

\newcommand{\nminustwospecial}{%
  Let~$G$ be a connected graph with $|V|$ vertices in which there are no three vertices~$a,b,u$ such that $u$ is $a$'s only neighbor and is also $b$'s only neighbor. Then under monotone preferences, there is always a connected \efouter\ allocation of $G$ for $n=|V|-2$ agents.
}
\begin{theorem}%
  \label{thm:|V|-2-special}
  \nminustwospecial
\end{theorem}

\begin{proof}
  Since \cref{obs:|V|-1} implies that the conclusion holds for all connected graphs with at most three vertices, we assume $G$ to be a graph with at least four vertices in which there are no three vertices~$a,b,u$ such that $u$ is $a$'s only neighbor and is also $b$'s only neighbor.
  We modify the picking procedure for $|V|-1$ agents behind \cref{obs:|V|-1} and show that it always computes a connected \efouter\ allocation.
  There are two phases.
  In phase one, we order the agents and let every one in the order pick his most preferred vertex among those still available after agents earlier in the order have picked.
  There remain two vertices, called $a$ and $b$, which we consider in the second phase.
  Without loss of generality, let us assume that the order of the agents is $1,\ldots,n$ with each agent~$i$ picking vertex~$v_i$ in phase one. %
  Note that no agent~$i$ values either $a$ or $b$ more than the vertex~$v_i$ that he currently has.
  In phase two, we distinguish between two cases:
  \begin{compactitem}[-]

    \item[Case I:]\label{twoneighbors}
    \newH{There exist vertices $v_i \neq v_j$ among $\{v_1,\ldots,v_n\}\setminus \{a,b\}$ with $v_i$ a neighbor of $a$ and $v_j$ a neighbor of $b$.}
Then we assign vertex~$a$ (along with $v_i$) to agent $i$ and assign~$b$ (along with $v_j$) to agent $j$, which maintains connectivity. As no agent has more than two vertices and the agents with two have either $a$ or $b$ (neither of which is more valuable to any agent than is the vertex that agent picked in phase one),
    the resulting allocation is \efouter.
  
    \item[Case II:]\label{notwoneighbors}
    \newH{Case I fails to hold.}
    Then as $G$ is connected, at least one of $a$ or $b$ must have a neighbor among $v_1,\ldots,v_n$, so without loss of generality assume that $b$ is adjacent to  $u \in \{ v_1,\ldots,v_n \}$.  As Case I fails to hold, $a$ is not adjacent to any vertex other than $u$ among $ v_1,\ldots,v_n $.  But vertex $a$ is not isolated, so it must be that either 
    \begin{compactenum}
    \item $a$'s sole neighbor is $b$, or else 
    \item $a$ has $b$ and $u$ as its only two neighbors.
    \end{compactenum}
    (Note that it is impossible for $a$'s sole neighbor to be $u$; if that happened, then the theorem's hypothesis---that $u$ not be both the sole neighbor of $a$ and the sole neighbor of $b$---would imply that $b$ must have a second neighbor among $v_1,\ldots,v_n$, putting us into Case~I, not II.)
    
    In either of these two sub-cases, we check whether some agent~$i$ finds the bundle~$\{a,b\}$ more valuable than his current vertex~$v_i$. If not, then we will give both $a$ and $b$ to the agent who already has $u$, which maintains the connectivity constraint.
    It is straightforward to verify that the resulting allocation is \efouter; note that for the agent that receives $\{u,a,b\}$ we can delete his picked vertex~$u$ to help establish \efouter.
    
    Otherwise, let $i$ be the first agent in the order $(1,\ldots, n)$ who finds $\{a,b\}$ more valuable than his current vertex~$v_i$.
    We remove $v_i$ from agent $i$'s share, giving him the piece~$\{a,b\}$ instead.
    Then, we let the agents from $\{i+1, \ldots, n\}$ (in the same order as originally used) pick single vertices from $\{v_i,\ldots,v_{n}\}$ in the same way as before, with each agent~$z$ in the order picking the most preferred vertex still available after all agents from $\{i+1,\ldots,z-1\}$ have picked.
    For all $z\in \{i+1, \ldots, n\}$, let $w_z$ denote the vertex picked by agent~$z$, and set $w_{z'}$ equal to $v_{z'}$ for all~$z'\in \{1, \ldots, i-1\}$; this defines the \emph{phase-two picking sequence}.
    Note that by using the original picking order, we know that each agent $j$ with $j \neq i$ either picks the same vertex in phase two that he had picked originally in phase one, or instead picks one that he prefers at least as much as the phase one choice.
    Now, there remains a last vertex; call it vertex~$x$. We'll show next that $x$ is adjacent to some vertex $w_j \in W = \{w_1, \ldots w_{i-1}, w_{i+1},\ldots w_n\} = V\setminus \{a,b,x\}$.

    In sub-case II(1), $b$ is the sole neighbor of $a$, so by the statements hypothesis, $b$ cannot also be the sole neighbor of $x$, whence $x$ must have some neighbor $w_j \in W$.

    In sub-case II(2), $a$'s only neighbors are $b$ and $u$. Also, $b$'s only neighbors are $a$ and $u$, for otherwise we would again be back in Case~I.  As $G$ is connected, $u$ must then be adjacent to some vertex $y$ other than $a$ or $b$, with $y$ equal to some $w_j \in W$.  Thus, if $x=u$, $x$ is adjacent to $w_j \in W$. On the other hand, if $x \neq u$, then as $G$ is connected, $x$ must have a neighbor $y$, which is not $a$ or $b$, so $y$ is equal to some $w_j \in W$.

In any of these three situations, we add $x$ to agent $j$'s share, with  $j$'s share becoming~$\{w_j,x\}$.
    The resulting allocation is connected.  We claim that it is also \efouter.
    Every agent $k$, except for agent $i$, had the option of choosing $x$ in the phase-two picking sequence, but instead chose an option $w_k$ they weakly prefer to $x$. So if we remove $w_j$ from agent $j$'s share $\{w_j, x\}$, agent $k$ does not envy agent $j$.  Any such agent $k$ also does not envy agent $i$, who receives share $\{a,b\}$, because in the phase-one sequence agent $k$ chose a vertex $v_k$ which they weakly preferred to either $a$ or $b$ alone, and in the phase-two sequence they received a vertex $w_k$ that they weakly preferred to $v_k$. Thus they feel no envy towards $\{a,b\}$ after either of these two vertices is removed; agent $j$ similarly feels no envy towards $\{a,b\}$ after either vertex of the two vertices is removed.  Agent $i$ had the option of choosing $x$ in the phase-one picking sequence (because we know that any agent $p$ with $p < i$ did not choose $x$ in either the phase-one or phase-two sequence) but instead chose $v_i$ and later replaced that choice by $\{a,b\}$, which they strictly preferred to $v_i$.  So, they strictly prefer their share to agent $j$'s share $\{w_j, x\}$, after $w_j$ is removed.  Of course, no agent feels envy towards an agent who received only one vertex, after that one is removed, and we conclude that the allocation is \efouter.     
  \end{compactitem}
  As we have constructed an \efouter\ allocation in both cases, the statement follows.
\end{proof}

So, what common patterns can we now assert for the spectra of \emph{all} graphs?
Each spectrum for a connected graph starts with a \emph{yes}, and is all \emph{yes} from the $(|V|-1)^{\text{\emph{th}}}$ position on. 
For many connected graphs, \cref{thm:|V|-2-special} tells us that they are all \emph{yes} already from the $(|V|-2)^{\text{\emph{th}}}$ position on.
Each spectrum with any \emph{no}s thus has a last \emph{no} at some location $j$. Accordingly, we record a \emph{YES} in capital letters at location $j+1$ to indicate that the sequence is all \emph{yes} from then on.
The following conjecture represents the simplest pattern that fits these two observations.%

\begin{conjecture}\label{conjecture-specture} The spectrum of any connected graph $G$ consists of an initial \emph{yes} string, followed by a (possibly empty) \emph{no} string, followed by an unending \emph{yes} string. 
\end{conjecture}

\noindent Conjecture \ref{conjecture-specture} is supported by several examples, as we see next. Note, also, that the middle \emph{no} string of the conjecture mirrors the ``critical interval'' in the Theorem \ref{thm:main} statement.
\newcommand{\lemspectrum}{
  \begin{compactenum}[(1)]
  \item The $5$-pointed star has spectrum $\langle${\emph yes, no, no, no, YES}$\rangle$.
  \item Graph IV in this paper, aka the \emph{friendly diamond} graph, has spectrum $\langle${\emph yes, yes, no, YES}$\rangle$.
  \item The version $L^{*}$ of the Lips graph found in \cite{IgZwi} has $8$ vertices (also see \cref{fig:VII+JCS+Lips-graph}), with spectrum  {$\langle${\emph yes, yes, yes, no, YES}$\rangle$}.
\end{compactenum}
}
\begin{lemma}%
  \label{lem:spectrum}
\lemspectrum
\end{lemma}

\begin{proof}
  A $5$-pointed star has a central vertex adjacent to $5$ vertices of degree $1$.  The three \emph{no}s in its spectrum follow from reasoning like that used for Graph I. %
  The \emph{YES} sits in the $5^{\text{th}}$ position because of the general rule (discussed above) for $n \ge |V|-1$.
  Similar reasoning applies to a $k$-pointed star, yielding a string of $k-2$ \emph{no}s.  For these graphs, all the \emph{yes} answers hold for arbitrary monotone valuations, and all \emph{no} answers arise from CA counterexamples, so the class $\mathcal{V}$ of valuations for this spectrum can be taken to be any of the classes of valuations that we have discussed.
  
  For Graph~IV with $6$ vertices in \cref{fig:gap2-examples}, the \emph{yes} in location $2$ follows from the bipolar ordering for the graph, and the \emph{no} at location~$3$ follows from the cutset of valence $2$ (both discussed earlier).  
  The \emph{yes} for $n\ge 4$ follows from \cref{thm:|V|-2-special} and \cref{obs:|V|-1}.   
  For the Lips graph~$L^{*}$ (see the right part of \cref{fig:VII+JCS+Lips-graph}), the second \emph{yes} uses the existence of a bipolar ordering paired with \cref{Bilo2Agents}.  The third \emph{yes}, proved in \cite{IgZwi}, uses the discretization (mentioned earlier) of a modified version of Stromquist's moving knife argument for three agents in \cite{Stromquist}.  The \emph{no} in the $4^{\emph th}$ position uses a graph cutset of valence $3$, consisting of the three vertices of degree greater than $2$ in~$L^{*}$.\footnote{That argument, from ``\emph {Example 1 continued}'' of \cite{IgZwi}, does not use the term ``graph cutset,'' but the idea is the same.}

  For five agents~$\{1,2,3,4,5\}$, we can modify the picking sequence strategy for the case with $n=|V|-2$ to obtain an \efouter\ allocation.
  First use the previous picking procedure to let the agents pick five of the eight vertices, \newH{called $v_1,\ldots,v_5,a,b,c$; note that we use the same names depicted in the right part of \cref{fig:VII+JCS+Lips-graph}.} Without loss of generality, let the picking order be $(1,\ldots, 5)$.
    There remain three vertices; call them $x,y,z$. Note that no agent prefers any vertex from $\{x,y,z\}$ over his current picked vertex.

  If these three vertices have three distinct neighbors that have been already picked, then we can allocate each of them to a separate agent without destroying the connectivity. The resulting allocation is \efouter\ since no agent receives more than two vertices and the agent that receives two vertices has one from $\{x,y,z\}$.

  If these three vertices do not have three distinct neighbors that have already been picked,then an exhaustive search reveals 9 possibilities for the set $\{x, y, z\}$, which we form into two groups as follows: $\{v_1, a, v_2\}, \{v_1, b, v_2\}, \{v_3, c, v_4\}$,  and $\{v_3, b, v_4\}$ are in Group $1$; with $\{a, v_1, b\}, \{a, v_2, b\},$ $\{c, v_3, b\}, \{c, v_4, b\}$, and $\{c, v_5, a\}$ in Group $2$. Each of these 9 sets has a path as induced subgraph, so each is connected and remains connected when either of the two endpoints of the path is removed. Let $\sigma(i)$ denote the vertex picked by agent~$i$, $i\in [5]$ initially.

  \begin{compactitem}[-]
    \item[Case 1:]
    Assume that $\{x,y,z\}$ belongs to Group $1$. The proof we give here for the special case $\{x, y, z\} = \{v_1, a, v_2\}$
    uses only that the middle vertex $a$ of the path is adjacent to some third vertex
other than the path's endpoints (here, $v_1$ and $v_2$), so a similar proof works for the other sets in Group $1$.
    We may need 2-3 rounds of reallocation.
    First, we check whether we can allocate $\{a, v_1\}$ to the agent that receives vertex~$v_5$ i.e., agent~$\sigma^{-1}(v_5)$, by checking whether for every agent~$i$ except agent~$\sigma^{-1}(v_5)$, it holds that he values $\sigma(i)$ at least as much as the worse of $\{a,v_1\}$ and $\{a, v_5\}$.
    If this is the case, then we can allocate $\{a,v_1\}$ to agent~$\sigma^{-1}(v_5)$ without violating the \efouter\ property.
    We allocate the remaining vertex~$v_2$ to agent
    $\sigma^{-1}(b)$. 
    This maintains \efouter.
  
    If someone (other than $\sigma^{-1}(v_5)$) likes the worse of $\{a,v_1\}$ and $\{a, v_5\}$ more than his current vertex, then let $i$ be the first agent in the order $(1, \ldots, 5)$ who prefers bundle~$\{a,v_1\}$ to his current vertex~$\sigma(i)$.
    We remove vertex~$\sigma(i)$ from agent~$i$ and allocate $\{a,v_1\}$ to him.

    Afterwards, we reallocate the vertices~$X=\{v_2,v_3,v_4,v_5,b,c\}$ to the agents except $i$ by allowing
    each agent in $(1,\dots,i-1, i+1,\dots,5)$ one-by-one pick his most preferred vertex from among those vertices in~$X$ that remain available to him.
      Let $x_k$ be the vertex that is picked by agent $k$ from $[5]\setminus \{i\}$.
      There remain two vertices.
      By our previous reasoning, one of them must be $v_2$.
      If the other one is not $b$, then we can assign each of the remaining vertices to a different agent and obtain an \efouter\ allocation.

      If the remaining vertices are $v_2$ and $b$, then we check whether some agent~$z$ from $[5]$ prefers 
      $\{b,v_2\}$ to his current share.
      We distinguish between three sub-cases:

      \begin{compactitem}[--]
      \item[Case 1(i):] If no agent from $[5]$ prefers his current share to $\{b,v_2\}$, then add $\{v_2,b\}$ to the share of the agent who picked $v_3$. It is straightforward to verify that the resulting allocation is \efouter.

      \item[Case 1(ii):] If no agent from $[5]\setminus \{i\}$ prefers $\{b,v_2\}$ to his current share~$\{x_{z}\}$, 
      but agent~$i$ prefers $\{b,v_2\}$ to his share $\{a,v_1\}$, then we switch the bundle of agent~$i$ from $\{a, v_1\}$ to $\{b,v_2\}$. 
      Next, allow the agents from $[5]\setminus \{i\}$ one-by one (in the original order, as always) to choose their preferred vertex from among those still available in $\{ a,c,v_1,v_3,v_4,v_5 \}$. 
      It must be that $a$ and $v_1$ are the two vertices that remain unpicked. Check whether it is possible -- without violating \efone -- to add $a$ and $v_1$ to the share of the agent who already holds $v_5$.  
      If so, do that, and we are done.  If not, let $j$ be the first agent in $[5]\setminus \{i\}$ who prefers $\{a,v_1\}$ to the single vertex $q$ that is her current share; note that from our previous reasoning, we know $j > i$. Remove $q$ and replace it with $\{a,v_1\}$ as agent $j$'s share. Then allow the remaining agents from $\{j+1, \dots, 5\}$ to choose anew from among those vertices that are not in the shares of agents $\{1, \dots, j\}$ and are still available.  There will be one unassigned vertex $y$, which is not among $a, v_1, v_2,$ or $b$. Any such $y$ has at least one neighbor $w$ other than $a, v_1, v_2,$ or $b$. Add $y$ to the share of the agent holding $w$. Three of the five shares now contain two vertices apiece, while the other two shares consist of a single vertex, and the allocation is \efouter.

      \item[Case 1(iii):] Some agent from $[5]\setminus \{i\}$ prefers $\{b,v_2\}$ to his current share~$\{x_{z}\}$. Then, let $j$ be the first agent in the picking sequence~$[5]\setminus \{i\}$ who prefers $\{v_2,b\}$ to his share,
      and assign $\{b,v_2\}$ as his share in place of ~$x_j$. Now have the remaining agents in $[5]\setminus \{i\}$ who come after $j$ in the order choose their preferred vertex from among those still available. As in the previous sub-case 1(ii) there will be one unassigned vertex $y$, which is not among $a, v_1, v_2,$ or $b$, and so the rest of the argument is the same as that for the previous sub-case.%
      \end{compactitem}

      \item[Case 2:] Assume that $\{x,y,z\}$ belongs to Group $2$. Four of the five sets in this group ($\{a, v_1, b\},$ $\{a, v_2, b\},$ $\{c, v_3, b\}$, and $ \{c, v_4, b\}$) are symmetric to one another, while the argument for the fifth set $\{c, v_5, a\}$ is quite similar to the others. So here we provide the argument for $\{x,y,z\} = \{a,v_1,b\}$. 
      
      First, check whether every agent other than~$\sigma^{-1}(v_2)$ likes her current share at least as much as the worse of  $\{a,v_1\}$ and $\{a, v_2\}$.  If so, we can can add  $a$ and $v_1$ to the share of agent ~$\sigma^{-1}(v_2)$, and add the single remaining vertex~$b$ to the share of agent~$\sigma^{-1}(v_3)$. The resulting allocation is \efouter. 
  If not, and some agent (other than~$\sigma^{-1}(v_2)$) prefers the worse of $\{a,v_1\}$ and $\{a,v_2\}$ to his current vertex,  let $i$ be the first such agent in the order $(1, \ldots, 5)$. The rest of this case follows just as in Case 1. (Agent $i$'s share now becomes $\{a,v_1\}$, we invite each agent in $(1,\dots,i-1, i+1,\dots,5)$ to re-pick from those vertices still available in $\{ b, c, v_2, v_3,v_4, v_5\}$. After that, one of the two remaining vertices must be $b$ while the second may or may not be $v_2$, and this is the same situation we had found ourselves in for the comparable stage of Case 1.)    

  \end{compactitem}
  
Finally, the \emph{yes} for $n \ge 6$ is due to \cref{thm:|V|-2-special} and \cref{obs:|V|-1}.  
 The class $\mathcal{V}$ for this spectrum can again be any of those we have mentioned.%
\end{proof}

\section{Conclusion}\label{sec:conclusion}
We have introduced graph \gencutset{s} as obstructions for the existence of connected \efone{} allocations and show that detecting such structure is NP-hard.
In terms of future research directions, it would be interesting to characterize important structures that preclude the non-traceable JCS graph from having a connected \efone\ allocation; such a characterization may have something interesting to tell us about  whether it is only traceable graphs that guarantee existence of connected \efone allocations for arbitrarily many agents (this was Question \ref{Q1}). 
Second, it would be interesting to look at the parameterized complexity of finding \gencutset{s}. Last but not least, it would be interesting to see whether \cref{conjecture-specture} holds true.

\section*{Acknowledgments}
Jiehua Chen is supported by the Vienna Science and Technology Fund (WWTF) [10.47379/ VRG18012].
The work has been partially supported by the Funds for International Relationships at TU Wien. 
William Zwicker was supported by the CY Initiative of Excellence (grant ``Investissements d’Avenir'' ANR-16-IDEX-0008).
We thank Manuel Sorge for suggestions that clarified some of these concepts. 

\todoHinline{Extensions. Parameterized complexity of finding generalized cutset.}

\bibliographystyle{abbrvnat} 
\bibliography{bib}

\end{document}